\documentclass[sigplan,10pt, nonacm]{acmart}
\renewcommand\footnotetextcopyrightpermission[1]{}
\usepackage{graphicx} 
\usepackage{subcaption}
\usepackage{tikz}
 
\usepackage{pifont}

\usepackage[T1]{fontenc}
\usepackage{beramono}
\usepackage{listings}
\usepackage{xcolor}
\usepackage{algpseudocode}
\usepackage{algorithm}
\usepackage{dsfont}
\usepackage{pgfplots}
\usepackage{pgfplotstable}
\usepackage{quantikz}
\usepgfplotslibrary{groupplots}
\usepackage{array,tabularx}
\pgfplotsset{compat=1.18}
\usepackage{tgheros}
\usetikzlibrary{arrows.meta,calc,shapes.geometric}

\definecolor{dkgreen}{rgb}{0,0.6,0}
\definecolor{gray}{rgb}{0.5,0.5,0.5}
\definecolor{mauve}{rgb}{0.58,0,0.82}

\newcommand{\as}[1]{\textcolor{blue}{as: #1}}

\lstdefinestyle{myScalastyle}{
  frame=tb,
  language=scala,
  aboveskip=3mm,
  belowskip=3mm,
  showstringspaces=false,
  columns=flexible,
  basicstyle={\small\ttfamily},
  numbers=none,
  numberstyle=\tiny\color{gray},
  keywordstyle=\color{blue},
  commentstyle=\color{dkgreen},
  stringstyle=\color{mauve},
  frame=single,
  breaklines=true,
  breakatwhitespace=true,
  tabsize=3,
}

\algblockdefx{AsyncLoop}{EndAsyncLoop}{\textbf{async loop}}{\textbf{end loop}}
\algblockdefx{Try}{EndTry}{\textbf{try}}{\textbf{end try}}
\algcblockdefx[Try]{Try}{Catch}{EndTry}[1]{\textbf{catch} #1}{\textbf{end try}}

\title{Qurator: Scheduling Hybrid Quantum Classical Workflows Across Heterogeneous Cloud Architectures}
\author{Sinan Pehlivanoglu}
\affiliation{%
  \institution{Indiana University Bloomington}
  \city{Bloomington}
  \state{Indiana}
  \country{USA}
}
\email{spehliva@iu.edu}

\author{Ulrik de Muelenaere}
\affiliation{%
  \institution{University of Notre Dame}
  \city{Notre Dame}
  \state{Indiana}
  \country{USA}
}
\email{udemuele@nd.edu}

\author{Peter Kogge}
\affiliation{%
  \institution{University of Notre Dame}
  \city{Notre Dame}
  \state{Indiana}
  \country{USA}
}
\email{kogge@nd.edu}

\author{Amr Sabry}
\affiliation{%
  \institution{Indiana University Bloomington}
  \city{Bloomington}
  \state{Indiana}
  \country{USA}
}
\email{sabry@iu.edu}

\begin{document}
\settopmatter{printfolios=true}
\pagestyle{plain}

\begin{abstract}
We present Qurator, a quantum-classical task scheduler that jointly optimizes
queue time and circuit fidelity across heterogeneous architectures. Qurator models
hybrid workloads as dynamic DAGs with explicit quantum semantics, including
entanglement dependencies, synchronization barriers, no-cloning constraints,
and circuit cutting and merging decisions. Fidelity is estimated through a unified
logarithmic success score that reconciles incompatible calibration data from
IBM, IonQ, IQM, Rigetti, AQT, Pasqal and other providers into a canonical set of gate error,
readout fidelity, and decoherence terms. We evaluate Qurator on a simulator
driven by four months of real queue data using circuits from the Munich
Quantum Toolkit benchmark suite. Across load conditions from 5 to 35,000
quantum tasks with up to 120,000 classical tasks, Qurator stays within 1\% of the highest-fidelity baseline at
low load, achieves 30--75\% queue time reduction at high load, all at a
fidelity cost bounded by a user-specified target while improving on list and fair schedulers as well as state of the art schedulers on both metrics. 
\end{abstract}

\maketitle
 
\section{Introduction}

Task scheduling plays a crucial role in achieving high
efficiency in computing systems. Large and complex compu-
tational workloads can be distributed and parallelized among
multiple processors in order to speed up the computation.
The objective of scheduling is to assign tasks to available
resources so as to minimize the \textit{makespan}, the total execution time of the tasks. The scheduling problem is considered
to be NP-hard in its general form, as well as in many
practical, restricted cases. Yet, its significance for complex
workflows in the age of big data and scientific computing has motivated extensive
research. Addition of quantum devices and quantum accelerated
workflows into these clusters introduces new degrees of
freedom and opportunities in scheduling, but also gives rise
to a new array of problems that limit existing classical task
scheduling techniques.

Quantum cloud providers like IBM, IonQ, and D-Wave expose shared Quantum Processing Units (QPUs)
through first-come first-serve HTTP queues. These systems are commonly referred to as \textit{loosely
coupled hybrid systems}.\footnote{We use ``hybrid'' exclusively in this sense
throughout, as opposed to the static-dynamic meaning common in scheduling
literature.}  Due to shared, public resources, a circuit that executes in
3 seconds can spend minutes to an entire day waiting: a 10-30,000$\times$
overhead~\cite{nation2024benchmarkingperformancequantumcomputing}.
Minimizing this overhead is not a straightforward application of
classical scheduling techniques. Quantum-specific constraints like entanglement
dependencies, the no-cloning theorem, non-preemptibility, heterogeneous
gate sets, and incompatible calibration metrics across providers require
fundamentally new scheduling semantics. Consequently, the addition of QPUs into heterogeneous systems, alongside GPUs and FPGAs, requires us to rethink task scheduling, and calls for a scheduler that can accommodate today's limitations, but also adapt to the rapidly evolving quantum road map. We present Qurator, an architecture-agnostic scheduler that jointly optimizes queue time and fidelity for loosely coupled quantum-classical workflows across heterogeneous providers. Qurator estimates device level lowering for each circuit and every candidate device and estimates its fidelity there from live calibration data and hardware specialized properties, treats circuit cutting and merging as dynamic decisions that not only depend on the circuit but the devices available, and jointly optimizes queue time and circuit fidelity across heterogeneous providers. It relies only on
information that providers expose publicly and without the need for long history or training, adapts quickly as devices are added and retired. Our main contributions are:

\begin{itemize}
    \item \textbf{Scheduler-oriented, unified fidelity estimation across heterogeneous
    architectures.} Each provider exposes incompatible calibration data.
    We reconcile gate error, readout fidelity, parallelism semantics and decoherence terms for superconducting, ion-trap and neutral atom devices provided by IBM, IonQ, IQM, Rigetti, AQT, and Pasqal into a canonical
    logarithmic success score, and validate it empirically against
    hardware outcomes across GHZ benchmarks of 2-10 qubits. 

    \item \textbf{End-to-end quantum-aware scheduling semantics.} Classical
    schedulers often leverage preemptibility, task duplication, work stealing and homogeneous
    task structure. We formalize the constraints that break these
    assumptions: entanglement synchronization barriers with a precise
    overhead measure, no-cloning restrictions on task duplication,
    circuit cutting and merging as first-class scheduling decisions,
    and dynamic DAG submission for iterative hybrid algorithms such as the Variational Quantum Eigensolver (VQE)  and the Quantum Approximate Optimization Algorithm (QAOA). We show that these semantics force the task-level properties to become scheduler-level concerns. Qurator is a scheduler designed to make practical, low-overhead decisions under incomplete information.

    \item \textbf{Circuit cutting \& circuit merging as scheduling strategies.} We heuristically deploy various circuit manipulation techniques such as circuit cutting and circuit merging, based on dynamic resource availability, instead of assuming a fixed circuit structure. Qurator dynamically optimizes the cuts and parallelization based on available devices and calibration data in order to generate more favorable schedules while allowing transformation subgraphs to be generated at run time and accounts for circuit generation, optimization, execution, verification, and cutting-result reconstruction in the end-to-end plan.
 
    \item \textbf{Load-adaptive joint optimization.} Qurator
    continuously rebalances fidelity and queue time as load grows and adjusts using a Gaussian kernel with submitted job metrics to estimate
    device wait times without access to queue internals. At low load
    it tracks the highest-fidelity baseline within 1\%; under high
    load (5,000--35,000 tasks) it achieves 30--75\% queue time
    reduction at a fidelity cost bounded by a user-specified target. For high qubit circuits, Qurator deploys circuit cutting to achieve up to 60\% gain in fidelity at the expense of added queue time for an increased number of tasks.

    \item \textbf{First scheduling benchmark for distributed entangled
    tasks.} Since today's public cloud does not support distributed
    quantum computing, no scheduling benchmark exists for entangled
    tasks. We extend and adapt Qurator to distributed execution semantics, introduce metrics for start skew, finish skew, budget
    penalty, and a survival proxy, and evaluate Qurator on a simulator
    driven by four months of real queue data across 11 devices and up
    to 35,000 quantum tasks from the Munich Quantum Toolkit (MQT) benchmark suite~\cite{quetschlich2023mqtbench}.
\end{itemize}

The rest of the paper is organized as follows. Sec.~\ref{sec:motivation} describes the hybrid task model and the
constraints that shape quantum cloud scheduling, and
Sec.~\ref{sec:fidest} presents the cross-architecture fidelity estimator.
Secs.~\ref{sec:cut} and~\ref{sec:merging} describe circuit cutting and
merging, and Sec.~\ref{sec:entangle} formalizes the synchronization barrier imposed by entangled tasks. Sec.~\ref{sec:architecture} describes the Qurator architecture and its scheduling algorithms,
Sec.~\ref{sec:evaluation} evaluates Qurator against multiple baselines, Sec.~\ref{sec:relatedwork} surveys related
work, and Sec.~\ref{sec:conc} concludes.


\begin{figure*}
    \centering
    \includegraphics[width=0.8\linewidth]{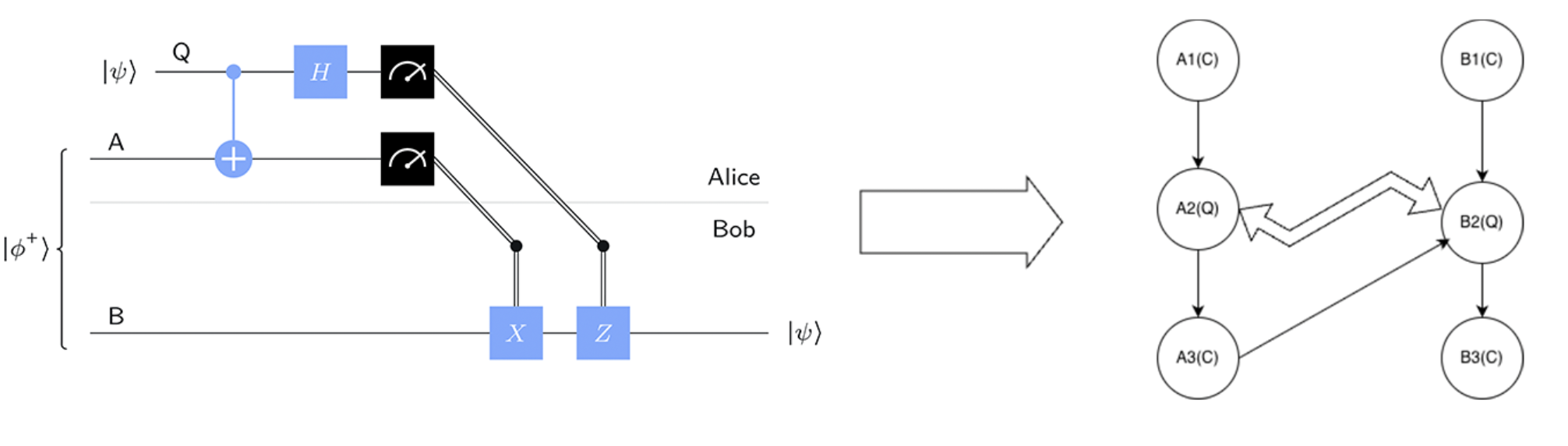}
    \caption{Quantum teleportation~\cite{qiskittel} as a hybrid workflow:
    the circuit (left) and its task DAG (right). Nodes labeled \textit{C}
    are classical tasks and nodes labeled \textit{Q} quantum tasks. After
    classical initialization, Alice's (A) and Bob's (B) quantum tasks share
    an entangled pair (double arrow); Alice's classical task A3 forwards her
    measurement outcome to Bob's quantum task B2.}
    \label{fig:qtel}
\end{figure*}

\begin{figure*}[htbp]
    \centering
    \resizebox{\textwidth}{!}{%
\begingroup
\fontfamily{qpl}\selectfont
\definecolor{qaInk}{HTML}{252A30}
\definecolor{qaMuted}{HTML}{606A74}
\definecolor{qaRule}{HTML}{9CA4AC}
\definecolor{qaAccent}{HTML}{375E7C}
\definecolor{qaFill}{HTML}{F4F6F7}
\tikzset{
  qaPanel/.style={draw=qaRule,line width=.45pt,fill=white},
  qaFlow/.style={draw=qaInk,line width=.8pt,-{Stealth[length=1.9mm,width=1.6mm]}},
  qaSync/.style={draw=qaAccent,line width=.75pt,dash pattern=on 3pt off 2pt,
    -{Stealth[length=1.7mm,width=1.4mm]}},
}
\newcommand{\qaText}[4]{\node[anchor=north west,align=left,text width=#3*.2mm,
  inner sep=0pt] at (#1,#2) {#4};}
\newcommand{\qaCenter}[4]{\node[anchor=north,align=center,text width=#3*.2mm,
  inner sep=0pt] at (#1,#2) {#4};}
\newcommand{\qaHeading}[5]{
  \qaText{#2}{#3+1}{26}{\color{qaMuted}\fontsize{8.5}{11}\selectfont #1}
  \qaText{#2+30}{#3}{#4}{\fontsize{10.3}{13.3}\selectfont\bfseries #5}}
\newcommand{\qaBullet}[4]{
  \fill[qaMuted] (#1+2,#2+8) circle[radius=1.6];
  \qaText{#1+15}{#2}{#3}{#4}}
\newcommand{\qaTask}[3]{
  \if C#1
    \draw[draw=qaMuted,line width=.6pt,fill=white] (#2,#3) circle[radius=15];
  \else
    \draw[draw=qaAccent,line width=.7pt,fill=qaAccent!9] (#2,#3) circle[radius=15];
  \fi
  \node[font=\fontsize{8.5}{9}\selectfont] at (#2,#3) {#1};}
\begin{tikzpicture}[x=.2mm,y=-.2mm,text=qaInk,
  font=\fontsize{8.5}{12.4}\selectfont]
\path[use as bounding box] (0,0) rectangle (1619,920);

\path[draw=qaRule,line width=.6pt,fill=white] (251,11) rectangle (1337,838);
\qaText{269}{23}{900}{\fontsize{12}{15}\selectfont\bfseries Qurator}

\path[qaPanel] (12,11) rectangle (229,838);
\qaHeading{1}{27}{29}{160}{Workflow\\submission}
\qaText{30}{108}{180}{Hybrid quantum--classical\\workloads represented\\as dynamic DAGs}
\draw[qaFlow] (107,230)--(69,260);
\draw[qaFlow] (129,230)--(155,257);
\draw[qaFlow] (55,285)--(55,317);
\draw[qaFlow] (61,347)--(76,382);
\draw[qaFlow] (83,412)--(83,440);
\draw[qaFlow] (170,285)--(178,317);
\draw[qaFlow] (178,348)--(162,382);
\draw[qaFlow] (121,348)--(145,386);
\draw[qaFlow] (155,412)--(155,440);
\draw[qaSync] (157,284)--(125,319);
\draw[qaFlow] (165,333)--(132,333);
\qaTask{C}{118}{217}
\qaTask{C}{55}{270}
\qaTask{Q}{166}{270}
\qaTask{Q}{55}{333}
\qaTask{Q}{116}{333}
\qaTask{C}{181}{333}
\qaTask{C}{83}{397}
\qaTask{Q}{155}{397}
\foreach \x in {83,155} \foreach \y in {454,460,466}
  \fill[qaMuted] (\x,\y) circle[radius=1.2];

\path[qaPanel] (266,78) rectangle (508,440);
\path[qaPanel] (532,78) rectangle (785,440);
\path[qaPanel] (807,78) rectangle (1059,440);
\path[qaPanel] (1084,78) rectangle (1318,440);
\qaHeading{2}{281}{94}{186}{DAG analysis}
\qaText{282}{141}{207}{Extract quantum semantics\\and structural features}
\draw[qaRule,line width=.35pt] (282,201)--(492,201);
\qaBullet{283}{219}{194}{Dependencies\\(including entanglement)}
\qaBullet{283}{273}{194}{Task runtimes}
\qaBullet{283}{307}{194}{Circuit properties}
\qaBullet{283}{341}{194}{Synchronization groups}
\qaBullet{283}{375}{194}{Resource demands}

\qaHeading{3}{547}{94}{198}{Resource state\\estimator}
\qaText{548}{163}{220}{Predict queue and execution\\state for all providers}
\path[draw=qaRule,line width=.4pt,fill=qaFill] (548,242) rectangle (769,309);
\qaText{560}{263}{202}{\bfseries Queue time prediction}
\path[draw=qaRule,line width=.4pt,fill=qaFill] (548,337) rectangle (769,404);
\qaText{560}{358}{202}{\bfseries Hardware state}

\qaHeading{4}{822}{94}{204}{Compilation, routing\\\& timed scheduling}
\qaText{823}{163}{220}{Transform circuits and map\\tasks to devices with\\temporal constraints}
\draw[qaRule,line width=.35pt] (823,241)--(1043,241);
\qaText{823}{253}{220}{\bfseries Circuit compilation}
\qaText{823}{277}{220}{\fontsize{7.5}{10}\selectfont Optimization, basis translation}
\draw[qaRule,line width=.35pt] (823,305)--(1043,305);
\qaText{823}{317}{220}{\bfseries Logical-to-physical routing}
\qaText{823}{341}{220}{\fontsize{7.5}{10}\selectfont Connectivity-aware mapping}
\draw[qaRule,line width=.35pt] (823,369)--(1043,369);
\qaText{823}{381}{220}{\bfseries Timed scheduling}
\qaText{823}{405}{220}{\fontsize{7.5}{10}\selectfont Dependencies and timing}

\qaHeading{5}{1099}{94}{191}{Fidelity estimator}
\qaText{1100}{141}{203}{Estimate circuit success\\probability on each provider}
\path[draw=qaRule,line width=.4pt,fill=qaFill] (1100,224) rectangle (1302,328);
\qaText{1111}{241}{182}{\fontsize{7.4}{10}\selectfont\bfseries\mbox{Logarithmic success score}}
\qaText{1111}{276}{182}{Calibration-derived}
\qaText{1100}{356}{202}{Unified across providers:\\IBM, IonQ, IQM, Rigetti,\\AQT, Pasqal}

\path[qaPanel] (266,467) rectangle (526,815);
\path[qaPanel] (551,467) rectangle (905,815);
\path[qaPanel] (935,467) rectangle (1248,815);
\qaHeading{6}{281}{484}{209}{Strategy generation}
\qaText{282}{533}{225}{Create candidate execution\\strategies}
\foreach \qaRowY in {592,636,680,724,768}
  \draw[qaRule,line width=.35pt] (282,\qaRowY-10)--(510,\qaRowY-10);
\qaText{288}{592}{214}{Placement}
\qaText{288}{636}{214}{Cutting}
\qaText{288}{680}{214}{Merging}
\qaText{288}{724}{214}{Batching}
\qaText{288}{768}{214}{Synchronized groups}

\qaHeading{7}{566}{484}{299}{Plan generation \& coordination}
\qaText{567}{533}{322}{Generate execution plans including\\cross-device synchronized groups}
\path[draw=qaAccent,line width=.6pt,dash pattern=on 3pt off 2pt,
  fill=qaAccent!3] (567,591) rectangle (889,735);
\qaText{581}{604}{296}{\color{qaAccent}\bfseries Synchronized execution}
\qaCenter{728}{642}{147}{\fontsize{7.8}{10.5}\selectfont Start-skew\\minimization}
\draw[qaAccent,dash pattern=on 3pt off 2pt,line width=.75pt] (671,685)--(785,685);
\path[draw=qaAccent,line width=.55pt,fill=white] (581,666) rectangle (671,704);
\path[draw=qaAccent,line width=.55pt,fill=white] (785,666) rectangle (875,704);
\node at (626,685) {Device A};
\node at (830,685) {Device B};
\qaCenter{728}{713}{295}{\fontsize{7.8}{10}\selectfont Synchronization barriers}
\qaText{567}{756}{322}{Plan includes: placement / cut / merge /\\batch / synchronized groups}

\qaHeading{8}{950}{484}{253}{Feasibility, synchronization\\\& plan scoring}
\qaText{951}{550}{281}{Evaluate and score plans}
\qaBullet{951}{592}{266}{Resource \& topology constraints}
\qaBullet{951}{625}{266}{Entanglement lifetime checks}
\qaBullet{951}{658}{266}{Synchronization overhead\\(start skew)}
\qaBullet{951}{711}{266}{Predicted Queue time}
\qaBullet{951}{744}{266}{Fidelity}
\qaBullet{951}{777}{266}{Composite plan score}

\path[qaPanel] (1360,11) rectangle (1610,359);
\qaText{1376}{29}{220}{\fontsize{10.3}{13}\selectfont\bfseries External data sources}
\draw[qaRule,line width=.4pt] (1376,65)--(1594,65);
\qaText{1376}{84}{220}{\bfseries Historical queue data}
\draw[qaRule,line width=.35pt] (1376,152)--(1594,152);
\qaText{1376}{171}{220}{\bfseries Hardware calibration\\\& topology}
\draw[qaRule,line width=.35pt] (1376,241)--(1594,241);
\qaText{1376}{260}{220}{\bfseries Provider APIs}
\qaText{1376}{290}{220}{IBM, IonQ, IQM, Rigetti,\\AQT, Pasqal}

\path[qaPanel] (1360,404) rectangle (1610,838);
\qaHeading{9}{1375}{423}{196}{Execution \&\\monitoring}
\qaText{1376}{472}{220}{Dispatch the selected plan\\and monitor progress}
\path[draw=qaRule,line width=.45pt,fill=qaFill] (1376,555) rectangle (1470,627);
\path[draw=qaRule,line width=.45pt,fill=qaFill] (1500,555) rectangle (1594,627);
\qaCenter{1423}{573}{91}{Quantum\\providers}
\qaCenter{1547}{573}{91}{Classical\\resources}
\draw[qaFlow,{Stealth[length=1.7mm,width=1.4mm]}-{Stealth[length=1.7mm,width=1.4mm]}]
  (1472,591)--(1498,591);
\path[draw=qaRule,line width=.45pt,fill=qaFill] (1376,697) rectangle (1594,805);
\qaText{1390}{715}{190}{\bfseries Feedback loop}
\qaText{1390}{748}{190}{Update coefficients with\\observed outcomes}

\draw[qaFlow] (229,235)--(251,235);
\draw[qaFlow] (508,235)--(532,235);
\draw[qaFlow] (785,235)--(807,235);
\draw[qaFlow] (1059,235)--(1084,235);
\draw[qaFlow,{Stealth[length=1.9mm,width=1.6mm]}-{Stealth[length=1.9mm,width=1.6mm]}]
  (1322,234)--(1360,234);
\draw[qaFlow] (526,637)--(551,637);
\draw[qaFlow] (905,637)--(935,637);
\draw[qaFlow] (1248,591)--(1360,591);
\draw[draw=qaInk,line width=.8pt,
  {Stealth[length=1.9mm,width=1.6mm]}-{Stealth[length=1.9mm,width=1.6mm]}]
  (387,815)--(387,854)--(1485,854)--(1485,805);

\end{tikzpicture}
\endgroup%
    }
    \caption{Qurator's Architecture}
    \label{fig:quarch}
\end{figure*}

\section{Scheduling in the Quantum Cloud}\label{sec:motivation}

Scheduling on hybrid quantum-classical systems is dynamic, non-preemptive,
heterogeneous scheduling on shared resources. In this section we present the  quantum background, the hybrid task model, the constraints that
break classical scheduling assumptions, and the resulting problem.

\subsection{Quantum Primer}

A \textit{qubit} can hold a \textit{superposition} of 0 and 1; measuring it
collapses its state to a classical bit. A \textit{quantum circuit} is a
sequence of \textit{gates} (unitary operations), possibly interleaved with
measurements. We use \textit{fidelity} in an operational sense: the
probability that a device produces a correct output.
Fidelity degrades through \textit{gate errors}, \textit{readout errors}
during measurement, and \textit{decoherence}, the gradual loss of quantum state caused by environmental noise. The \textit{No-Cloning Theorem} states that an arbitrary, unknown quantum state cannot be copied.

Two qubits are \textit{entangled} when their joint state cannot be written
as a product of single-qubit states, so measurements on them give
correlated outcomes even when the qubits are far apart. Entangled pairs,
or \textit{EPR pairs}, are the basic resource of distributed quantum
computing. Once distributed to separate devices over fiber and quantum
repeaters, they survive for only a few seconds on current networks, so the
tasks that share a pair must start almost simultaneously.

\subsection{Hybrid Task Model}

We model a hybrid workflow as an annotated \textit{directed acyclic graph}
(DAG) $G = (V, E)$ whose vertices are classical or quantum tasks and whose
edges are dependencies; a task can start only when all its predecessors
have completed. Fig.~\ref{fig:qtel} shows the DAG for quantum teleportation,
which combines classical dependencies, quantum tasks, and an entanglement
relation between tasks on different devices.

Crucially, the DAG is not known ahead of time. Quantum tasks are generated
classically, from parameters commonly available  at run time. In variational
algorithms such as the Variational Quantum Eigensolver (VQE) and the
Quantum Approximate Optimization Algorithm (QAOA), a classical optimizer
updates parameters from measurement outcomes and emits a fresh layer of
circuits at each iteration. Neither the circuits nor the number of
iterations can be determined statically, which rules out static scheduling
and any dynamic technique that assumes prior knowledge of the task graph.

The DAG does not capture the possibly nontrivial cost of launching a
quantum task~\cite{10313739}, therefore the host must parse the circuit, initialize
the QPU, and configure the control hardware before execution begins, a
process analogous to kernel launch in GPU scheduling but considerably more
expensive.

\subsection{Device Constraints}

Today's \textit{noisy intermediate-scale quantum} (NISQ) devices impose
four constraints that shape scheduling.

\textbf{Few, sparsely connected qubits.} Devices offer at most a few
hundred qubits, and on superconducting and neutral-atom devices two qubits
must be adjacent to interact, so a circuit's interactions must be routed
through the device topology at the cost of extra gates and depth. When no
available device can run a circuit at the target fidelity, \textit{circuit
cutting}~\cite{tangCutQCUsingSmall2021} partitions it into smaller
subcircuits whose results are recombined by classical post-processing.
Smaller subcircuits also suffer less gate error and decoherence, so cutting
trades classical overhead for fidelity.

\textbf{Short coherence.} A qubit keeps its state only for a
device-specific coherence time, from about $10^{-4}$\,s on superconducting
devices to around a second on some trapped-ion devices. This bounds the
depth a circuit can reach reliably, and, unlike in classical computing, a
slower execution costs fidelity as well as time. Jobs themselves are short,
under 5 seconds on average on publicly accessible devices, so queueing
rather than execution dominates turnaround and a scheduling decision
that takes longer than the job’s execution time is self-
defeating.

\textbf{Heterogeneous gate sets and semantics.} Providers differ in native
gates, connectivity, parallelism, and measurement behavior, much as
classical processors differ in instruction sets. Because the target device
is known only at run time, the scheduler must \textit{transpile} each
circuit for every candidate device and account for the transpilation time.
Sec.~\ref{sec:fidest} shows how much the result can differ across devices.

\textbf{Shared, non-preemptible devices.} The cost and cooling requirements
of quantum hardware preclude dedicated or virtualized access. Users share
each device through a provider-managed queue, and a job cannot be
preempted once it starts running. Batch sessions, where offered, still
enter the public queue, and the circuits within them remain
non-preemptible. Running several circuits on one QPU therefore requires
merging them into a single submission. Merging does not guarantee parallel execution.
Devices that execute gates serially, such as trapped-ion systems with a
single interaction zone, still run the merged gates one after another, and
a merged submission completes only when its slowest constituent does.
Merged circuits also compete for the best qubits and gates, so the
scheduler must weigh device topology rather than maximize utilization.

These limitations break several classical scheduling assumptions. The
No-Cloning Theorem rules out work stealing and task duplication based scheduling techniques in the distributed setting. A
quantum state cannot be copied, so a cross-entangled task cannot be duplicated
without simultaneously duplicating and imposing a significant EPR generation overhead. Even though tasks solely consuming classical input can be trivially duplicated or stolen, tasks consuming entangled states or live quantum states such as magic states impose a significant penalty for doing so. While this may not immediately be interpreted as a problem for today's hybrid VQE/QAOA tasks, various techniques such as circuit cutting with communication channel assumptions might introduce entanglement without the programmer being aware of it. Therefore, scheduling techniques that rely on work stealing or task duplication do not generalize. Non-preemptibility eliminates time-sharing, therefore running multiple circuits
on a single QPU must instead be achieved by merging them into a single
submission. Whether this constitutes true parallelism depends on the device
architecture. 

\begin{figure}
    \centering
    \includegraphics[width=0.9\linewidth]{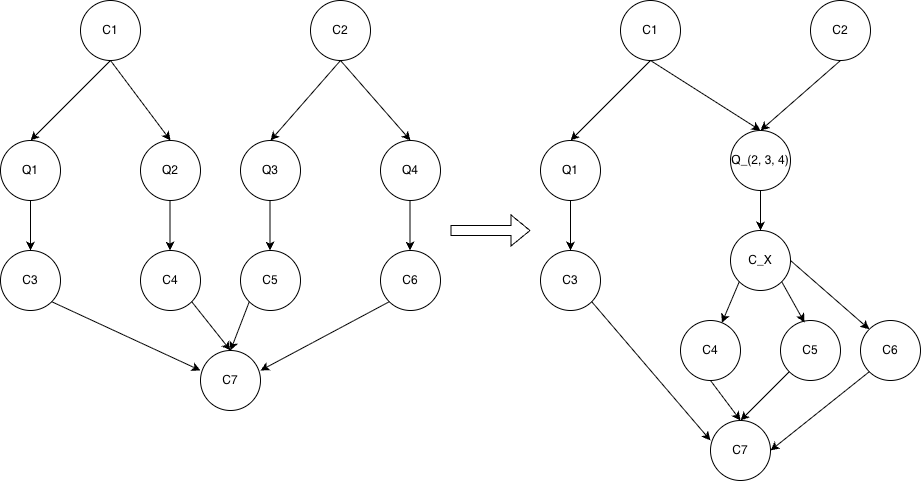}
    \caption{Example DAG transformation for merging quantum tasks to execute them in parallel on a single QPU.}
    \label{fig:quantparDAG}
\end{figure}

\subsection{The Scheduling Problem}

Scheduling a quantum task therefore pursues two competing objectives:
running the circuit on a device that achieves sufficient fidelity, and
minimizing makespan, which on the public cloud is dominated by queue time.
Existing approaches often optimize one at the expense of the other. The scheduler must also decide with partial
information. Providers differ in
whether they expose queue length, job cancellation, advance reservations,
and device topology, and none exposes per-job characteristics, so Qurator
relies on queue lengths, calibration snapshots, and historical traces.

\begin{figure}
    \centering
    \includegraphics[width=\columnwidth]{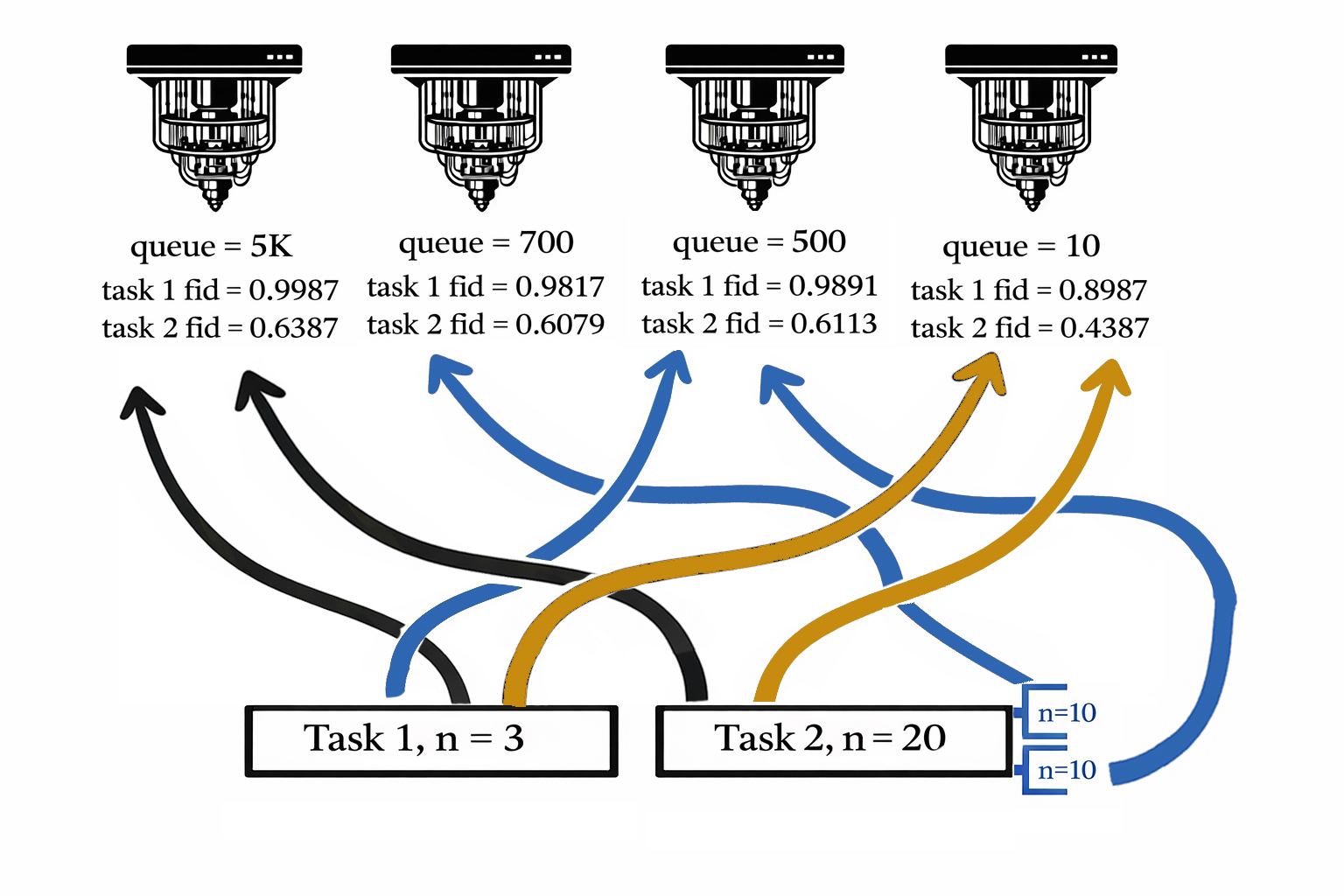}
    \caption{Scheduling a 3-qubit and a 20-qubit task with the
    \textit{least busy} (gold) and \textit{highest fidelity} (black)
    strategies and with Qurator (blue). Each device lists its queue length
    and the estimated fidelity (fid) of each task on it; $n$ is the number
    of qubits. Outcome fidelities are provided for uncut circuits.}
    \label{fig:schedex}
\end{figure}

Qurator resolves the tension device by device. It ranks the devices by a score that weighs estimated
fidelity against normalized queue length, shifting weight toward queue
length as load grows. Cutting and merging
change the candidates themselves: a cut circuit can run on smaller, less
loaded devices, and a merged submission takes one place in a queue instead
of several. Fig.~\ref{fig:schedex} illustrates the fidelity-queue tension for a small
3-qubit task and a larger 20-qubit task. The \textit{high fidelity} strategy
selects the best device for each circuit, but that device is heavily loaded
and both tasks join a 5,000-task queue. The \textit{least busy} strategy
selects a device with a $500\times$ shorter queue, but at a cost of 10--20
percentage points of fidelity. For the larger task, the least busy device also
lacks sufficient qubits, making those results effectively unusable. Qurator
navigates this tradeoff by selectively applying circuit cutting to the larger task,
then restricting device selection to those whose estimated fidelity exceeds a
user-specified target, and finally minimizing queue time within that set. The
result is fidelity close to the high-fidelity strategy for both tasks, with an
order-of-magnitude reduction in queue length. Qurator's architecture is summarized in Fig. \ref{fig:quarch}. We detail these steps in Sections \ref{sec:fidest}-\ref{sec:architecture}.


\begin{table*}
    \centering
    \begin{tabular}{c|p{15cm}}
       Provider & Metrics \\ \hline 
       IBM & T1, T2, Readout assignment error, Prob meas0 prep1, Prob meas1 prep0, Readout length, ID error, Single-qubit gate length, RX error, Pauli X error, CZ error, Gate length (ns), RZZ error, MEASURE error, MEASURE2 error \\ \hline 
       IonQ \& IQM & T1, T2, Average 1Q fidelity, Average 2Q fidelity, Average readout fidelity, 1Q gate duration, 2Q gate duration, Readout duration \\  \hline 
       Rigetti & T1, T2, Average 1Q fidelity, Readout Fidelity, Swap Fidelity, CZ Fidelity \\ \hline 
       QuEra & Absolute position error, Site position error, Atom position error, Typical filling error, Worst filling error, Typical vacancy error, Worst vacancy error, Typical atom loss probability, Worst atom loss probability, Typical atom capture probability, Worst atom capture probability, Typical atom detection false positive error, Worst atom detection false positive error, Typical atom detection false negative error, Worst atom detection false negative error \\ \hline
       Pasqal & State Prep Error, P False Positive, P False Negative, Relaxation Rate, Dephasing Rate, Laser Waist, Rydberg Level, Atom Distance, Atom Number, Trap Number, Interaction Coefficient, Gate Noise Channels \\
    \end{tabular}
    \caption{Calibration Metrics Reported By Different Providers as of January 2026}
    \label{tab:fidelity}
\end{table*}

\section{Estimating Fidelity}\label{sec:fidest}

Fidelity is not a property of a circuit alone. Each backend lowers a
logical circuit through a chain of device-specific steps,
\[
  \begin{gathered}
    \text{logical circuit} \rightarrow \text{native circuit} \rightarrow
    \text{routed circuit} \\
    \rightarrow \text{timed schedule} \rightarrow \text{pulse program}
  \end{gathered}
\]
each of which can change gate count, depth, parallelism, idle time,
measurement behavior, and noise exposure. Two devices that accept the same
circuit can therefore execute it very differently. A circuit with high
logical parallelism may run efficiently on one architecture and serialize
on another, and one with poor connectivity may need heavy routing on a
superconducting device but none on a trapped-ion device. Measurement crosstalk forces all sub-circuits to complete before any can be measured, so the runtime of a merged  batch equals that of the slowest constituent. Furthermore, parallelism increases competition for the highest-fidelity qubits and gates, requiring the scheduler to account for device topology and refrain from maximum utilization. A scheduler that
treats QPUs as interchangeable circuit executors will make systematically
wrong placement and batching decisions. Prior schedulers typically model
only connectivity or gate-set compatibility.
Qurator instead compiles, routes, and times each circuit for every
candidate device, whether superconducting, trapped-ion, or neutral-atom, and estimates fidelity on the result, so the programmer need not understand each architecture's physical trade-offs.

Two obstacles complicate this on the cloud. Providers report different and
incompatible calibration metrics (Table~\ref{tab:fidelity}), and none
exposes the device-level schedule before a circuit runs; IBM, which offered
pulse-level programming, recently retired this capability. Qurator therefore reconciles
provider data into a canonical set of quantities and reconstructs the timed schedule from device
properties.

\subsection{The Estimator}\label{sec:estimator}

We express every provider's calibration data in terms of the following
canonical quantities: one-qubit gate error $\varepsilon_{1q}$, two-qubit
gate error $\varepsilon_{2q}$, readout error $\varepsilon_{ro}$, one- and
two-qubit gate durations $\tau_{1q}$ and $\tau_{2q}$, readout duration
$\tau_{ro}$, and coherence times $T_1$ and $T_2$. For a circuit compiled,
routed, and scheduled for device $d$, define the operational success score
$P_{\mathrm{ops}}$ as:
\[
\begin{array}{rl}
\log P_{\mathrm{ops}}
=&
\displaystyle\sum_{(g,q)\in G_{1q}}
\log\bigl(1-\varepsilon_{1q}(g,q)\bigr)
\\[8pt]
+&
\displaystyle\sum_{(g,e)\in G_{2q}}
\log\bigl(1-\varepsilon_{2q}(g,e)\bigr)
\\[8pt]
+&
\displaystyle\sum_{q\in M}\log\bigl(1 - \varepsilon_{ro}(q)\bigr),
\end{array}
\]
where $G_{1q}$ is the set of realized one-qubit gates, $G_{2q}$ is the set of
realized two-qubit gates, and $M$ is the set of measured qubits. When
asymmetric readout errors are provided, $\varepsilon_{ro}(q)$ is reconstructed
as $(\Pr(0\mid 1;q)+\Pr(1\mid 0;q))/2$. Let $t_q$ denote the elapsed time from the
initialization of physical qubit $q$ until its final use or
measurement, including idle intervals in the timed schedule; the durations
$\tau$ enter the estimate through this schedule. We additionally define a decoherence penalty
$P_{\mathrm{decoh}}$ as
\[
\log P_{\mathrm{decoh}}
\approx
\sum_q -\frac{t_q}{T_{\mathrm{coh}}(q)}, \quad T_{coh}(q) = (\frac{1}{6T_1(q)} + \frac{1}{3T_2(q)})^{-1},
\]
 where $T_{\mathrm{coh}}$ is the provider-specific coherence
quantity used by the estimator. The combined estimated success probability is then
\[
\widehat{F}_{\mathrm{est}}
=
\exp\bigl(\log P_{\mathrm{ops}} + \log P_{\mathrm{decoh}} + \log P_{xt} + \log P_{\mathrm{trans}} + \log P_{\mathrm{init}}),
\]
where $\log P_{\mathrm{init}}$ and $\log P_{\mathrm{trans}}$ are optional values that are estimated from provider metrics such as \textit{prep errors, vacancy error, filling error, atom loss probability} errors and \textit{transport error, move/reload survival} as appropriate. Crosstalk factor $\log P_{xt}$, detailed in the next section, is estimated where appropriate but other correlated noise, leakage, or non-Markovian effects are not accounted for. This model is
intended for ranking devices and filtering them against a target, for
which a cheap estimate that orders devices correctly matters more than
point accuracy.

\subsection{Device-Specific Lowering}\label{sec:hardware}

Each lowering step feeds a specific term of the estimate. Native-basis
decomposition and placement determine which gates are realized and on
which qubits and couplers, and hence which calibrated errors enter
$\log P_{\mathrm{ops}}$: the ``same'' gate has a different error on
different qubits. Routing adds gates to $\log P_{\mathrm{ops}}$ and depth
to the schedule, and together with parallelism, directional gates and idle time it determines
each $t_q$, and hence $\log P_{\mathrm{decoh}}$. Routing or serialization
inserted for one group of qubits lengthens $t_q$ for every qubit waiting on it. The same circuit may display different SWAP overhead on two different devices depending on how well its two-qubit interactions embed naturally into the backend topology. Routing overhead significantly increases circuit depth and increases the fidelity loss. Parallel-gate support determines how much of the circuit DAG can actually be compressed into physical layers, while crosstalk, frequency collisions, gate-zone resources and multiplexed readout impose backend-specific constraints on which apparently independent operations can safely run simultaneously. Table~\ref{tab:lowering} summarizes how each architecture shapes these
steps.

The same work can take very different times on different devices. A
superconducting device with 20\,ns gates and threefold parallelism executes
1,000 gates in about $6.7\,\mu s$, while a trapped-ion device with
$1\,\mu s$ gates and no parallelism needs $100\,\mu s$ for 100 gates, 15
times longer for a tenth of the work performed. Duration alone does not decide
fidelity. However, it should be noted that $\log P_{\mathrm{decoh}}$ charges each qubit's time
relative to its coherence time, and trapped-ion coherence times are commonly orders
of magnitude longer. 

Two of the lowering steps are invisible from outside the device.
Parallelism is not specified by the programmer but generated by the control
system along with the timed schedule, and providers reveal that schedule
only after a circuit has run, so Qurator estimates the achievable
parallelism at each time step from the device properties providers expose.
Some providers, such as IonQ, do not expose their compiler at all, but they accept pre-compiled circuits, which lets Qurator submit the native circuit it scored. Qurator's modular and flexible compiler interface is therefore aware of directional and virtual gates, gate durations, routing, and atom placement, taking into consideration the unique capabilities and limitations of each device family shown in Table~\ref{tab:lowering}.  The same effects decide whether cutting or merging
pays off on a given device, so Qurator makes those decisions on the
compiled, device-specific schedules.

Qurator estimates parallelism on per-architecture basis, using device topology and control overlap for superconducting devices, gate-zone assignments and laser/transport-channel capacities for ion traps; and atom positions, exclusion radii, and addressing capacity for neutral atoms. Ion-trap gates \(i,j\) may overlap only if \(Z_i\cap Z_j=\varnothing\), where \(Z_i\) denotes the gate zones occupied by gate \(i\); simultaneous gates and transport operations are additionally bounded by their respective channel capacities. For disjoint neutral-atom gates, overlap requires 
\[
\min_{a\in Q_i,\;b\in Q_j}
\|\mathbf r_a-\mathbf r_b\|
>
R_{ij}+2\sigma,
\qquad
N_{\mathrm{active}}\le C_{\mathrm{addr}},
\]
where \(Q_i\) is the set of atoms used by gate \(i\), \(\mathbf r_a\) is atom \(a\)’s position, \(\sigma\) is the supplied positional uncertainty margin, and \(C_{\mathrm{addr}}\) is the maximum number of independently addressable simultaneous gates. The exclusion distance is
\[
R_{ij}=
\begin{cases}
R_{\mathrm{addr}}, & \text{both gates are single-qubit},\\
\max(R_b,R_{\mathrm{addr}}), & \text{otherwise},
\end{cases}
\]
with \(R_b\) the Rydberg blockade radius and \(R_{\mathrm{addr}}\) the addressing exclusion radius that is calculated using the provider exposed (i.e. Pasqal) interaction coefficients. Qurator's parallelism estimation directly impacts fidelity through $t_q$ and cross-talk noise $\log P_{xt}$. The former is directly determined by the constructed schedule depth, while the latter is estimated as sum over time slices. For a gate $i$ that occupies the pulse interval $[s_i, e_i)$ with $\tau_i = e_i - s_i \geq 0$ and for each unordered pair of overlapping gates, define: 
\[
  \omega_{ij} = \frac{\max(0, \min(e_i, e_j) - \max(s_i, s_j))}{\min(\tau_i, \tau_j)},
\]
\[
  \log P_{xt} = \sum_{\{i,j\} \in \mathcal{C}}\omega_{ij} \log S_{ij},
\]
where $\mathcal{C}$ contains each eligible gate pair once and \(S_{ij}\) is its modeled survival factor under full overlap, derived from the provider specific metrics covered in Table \ref{tab:fidelity} such as parallel benchmarks; optical waist, position uncertainty, crosstalk amplitude over the set of atoms used by the corresponding gates. 

\begin{table*}[t]
    \centering
    \footnotesize
    \setlength{\tabcolsep}{4pt}
    \begin{tabular}{@{}>{\raggedright\arraybackslash}p{0.12\textwidth}>{\raggedright\arraybackslash}p{0.27\textwidth}>{\raggedright\arraybackslash}p{0.27\textwidth}>{\raggedright\arraybackslash}p{0.27\textwidth}@{}}
    & \textbf{Superconducting} & \textbf{Trapped ion} & \textbf{Neutral atom (gate model)} \\ \hline
    \textbf{Native gates and routing} ($\log P_{\mathrm{ops}}$, $t_q$, $\log P_{\mathrm{init}}$, $\log P_{\mathrm{trans}}$) &
    Sparse coupling graph: non-adjacent interactions need SWAPs, and the overhead depends on how well the circuit's interaction graph embeds in the topology. Some two-qubit gates are directional; reversing one adds single-qubit basis changes. Gates such as RZ are virtual frame updates with essentially zero duration and error, and must not be counted as pulses. &
    Typically all-to-all connectivity: no routing, at the price of longer gates. In transport-based (QCCD) designs, moving ions adds to schedule length. &
    Compilation is driven by atom placement in a 2D geometry; interaction patterns that do not embed require atom movement or reconfiguration, which changes cost and depth. \\ \hline
    \textbf{Parallelism} ($t_q$, $\log P_{xt}$) &
    Gates on disjoint qubits can run in parallel, but crosstalk, frequency collisions, and readout crosstalk limit which can safely do so. &
    Device-dependent: IonQ devices generally do not parallelize operations; QCCD devices group gates into layers in gate zones with each ion participating in a single gate per layer, respecting dependency/order constraints, but splitting, merging, shuttling, and cooling can dominate schedule length. &
    Limited by the Rydberg blockade radius and local-addressing crosstalk: operations on disjoint atoms can still conflict if they are too close. Local addressing crosstalk means that parallelism cannot be inferred solely from disjoint qubit sets and can corrupt supposedly independent gates and introduce additional errors.  \\ \hline
    \textbf{Mid-circuit measurement} ($t_q$, qubit availability, gate count) &
    Support varies by device. &
    Can leave a qubit outside the computational subspace or add crosstalk. Because transport is not conditional, a measurement-conditioned branch can cost the depth of all branches. &
    Can lose atoms, requiring reload or rearrangement before later tasks. \\
    \end{tabular}
    \caption{How each architecture shapes the lowering steps, and which estimator inputs each step affects.}
    \label{tab:lowering}
\end{table*}

\subsection{Validation}\label{sec:validation}

To compare estimates with hardware, we use an empirical success
probability: given $N$ shots with outputs $x_1,\dots,x_N$ and ideal support
set $\mathcal{S}_{\mathrm{ideal}}$,
\[
\widehat{F}_{\mathrm{actual}}
=
\frac{1}{N}\sum_{i=1}^{N}\mathbf{1}\{x_i\in \mathcal{S}_{\mathrm{ideal}}\}.
\]
For an $n$-qubit GHZ circuit, $\mathcal{S}_{\mathrm{ideal}}=\{0^n,1^n\}$.
Fig.~\ref{fig:single-provider-fidelity} compares estimated and measured
success probabilities for GHZ circuits of 2, 4, 7, and 10 qubits on all six
providers; sizes are capped at 10 qubits so that every size fits the
smallest device, AQT's 12-qubit machine.

The estimate is within 0.05 of the measurement in 20 of 24 cases with a mean
absolute error 0.035. At 2, 4, and 10 qubits it overestimates on every
provider, by 0.015--0.026 on average. The large errors are at 7 qubits,
where it underestimates the two trapped-ion devices, IonQ by 0.10, AQT by
0.20, and overestimates Rigetti by 0.09. Because the scheduler uses the
estimate to rank devices, rank agreement matters more than absolute error.
The estimated and measured orderings agree throughout the experiments with at most one out-of-place candidate. At every size, the estimate identifies the best devices as the hardware.

 We additionally compared IBM noisy device snapshot simulators against our fidelity estimator, observed fidelity on real quantum hardware and GHZ fidelity through tomography for 2-10 qubit GHZ circuits as well as noisy Pauli simulator. We implement a Monte Carlo Pauli-frame evaluator using established Clifford error-propagation and Pauli-twirling methods \cite{Gidney2021stimfaststabilizer, Geller_2013}. For non-Clifford gates, we approximate propagated errors by sampling squared Pauli-expansion coefficients, discarding coherent interference \cite{Nielsen_2002}. The results are presented in Fig. \ref{fig:ibm_snapshotfidelity}. We additionally verify the Pauli simulator implementation on Clifford MQT circuits against Stim v1.16.0 \cite{Gidney2021stimfaststabilizer} with 1,000,000 samples per point with shared seeds, calibration and timing as our Pauli implementation, and observe a mean absolute difference of $0.000767$. We observe that noisy snapshot simulation overestimates device fidelity, while trace-based simulation falls slightly below real device execution and closely follows GHZ tomography results. We observe that the deviation of the trace-based simulator from the real hardware results remains close to that of noisy simulators and at times outperforms it while offering a uniform approach across all hardware. Additionally, underestimation still retains correctness as opposed to overestimating snapshots, further motivating use of this method over noisy snapshots. 
 
More accurate, architecture-specific estimators have been
proposed~\cite{Wang_2022}, but they require substantial training data,
incur higher computational overhead, and do not readily generalize to an
architecture-agnostic, multi-provider setting. For scheduling, a cheap
estimator that orders devices correctly is the better trade. Validation
on larger circuits is left to future work.

\begin{figure}[t]
\centering
\begin{tikzpicture}
\begin{axis}[
    width=\linewidth,
    height=0.42\textwidth,
    ymin=0, ymax=1.05,
    ylabel={Fidelity / success probability},
    xlabel={Provider},
    symbolic x coords={IonQ,AQT,IQM,Rigetti,Pasqal, IBM},
    xtick=data,
    x tick label style={rotate=20, anchor=east, font=\small},
    ytick={0,0.2,0.4,0.6,0.8,1.0},
    grid=both,
    legend columns=4,
    legend style={
        at={(0.5,1.18)},
        anchor=south,
        draw=none,
        fill=none,
        font=\small
    },
    tick label style={font=\small},
    label style={font=\small},
    title style={font=\small},
]

\addplot+[mark=o, thick] coordinates {
      (IonQ,0.96) (AQT,0.88) (IQM,0.89) (Rigetti,0.875) (Pasqal,0.93) (IBM,0.9633) 
};
\addlegendentry{Actual 2q}

\addplot+[mark=square*, thick] coordinates {
    (IonQ,0.9246) (AQT,0.8272) (IQM,0.8737) (Rigetti,0.835) (Pasqal,0.8912) (IBM,0.9423)
};
\addlegendentry{Actual 4q}

\addplot+[mark=triangle*, thick] coordinates {
    (IonQ,0.723) (AQT,0.66) (IQM,0.6872) (Rigetti,0.630) (Pasqal,0.7996) (IBM,0.8549)
};
\addlegendentry{Actual 7q}

\addplot+[mark=diamond*, thick] coordinates {
    (IonQ, 0.6812) (AQT,0.37) (IQM,0.5986) (Rigetti,0.614) (Pasqal,0.6553) (IBM,0.6998)
};
\addlegendentry{Actual 10q}

\addplot+[mark=o, thick, dashed] coordinates {
   (IonQ,0.973) (AQT,0.8977) (IQM,0.9123) (Rigetti,0.9247) (Pasqal, 0.9589) (IBM,0.9877)
};
\addlegendentry{Est. 2q}

\addplot+[mark=square*, thick, dashed] coordinates {
    (IonQ,0.9332) (AQT,0.8333) (IQM,0.8828) (Rigetti,0.8814) (Pasqal,0.9012) (IBM,0.9534)
};
\addlegendentry{Est. 4q}

\addplot+[mark=triangle*, thick, dashed] coordinates {
    (IonQ,0.6181) (AQT,0.46) (IQM,0.7122) (Rigetti,0.72) (Pasqal,0.8151) (IBM,0.8698)
};
\addlegendentry{Est. 7q}

\addplot+[mark=diamond*, thick, dashed] coordinates {
    (IonQ,0.6902) (AQT,0.4326) (IQM,0.6002) (Rigetti,0.6434) (Pasqal,0.6676)(IBM,0.7199)
};
\addlegendentry{Est. 10q}

\end{axis}
\end{tikzpicture}
\caption{Measured (solid) and estimated (dashed) success probability of GHZ circuits of 2--10 qubits on each provider; the marker encodes circuit size.}
\label{fig:single-provider-fidelity}
\end{figure}

\begin{figure}
    \centering
    \includegraphics[width=0.5\linewidth]{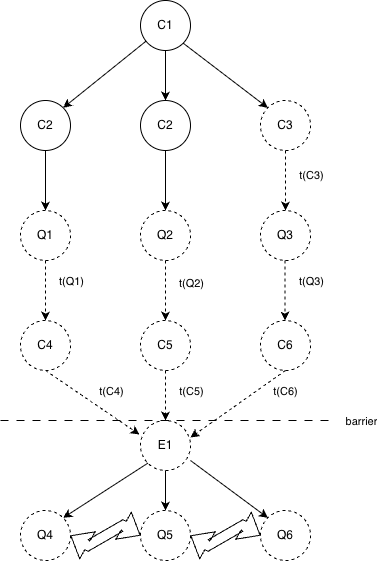}
    \caption{Synchronization barrier imposed by entanglement distribution.
    Solid nodes represent tasks known at compile time; dashed nodes are
    submitted dynamically at run time. $E_1$ cannot start until all three
    upstream paths complete. Once $E_1$ finishes, it distributes EPR pairs
    to $Q_4$, $Q_5$, and $Q_6$, which execute on separate devices and must
    begin as close together in time as possible before the pairs decohere.}
    \label{fig:entanglementsync}
\end{figure}


\begin{figure*}[t]
\centering
\begin{tikzpicture}

\begin{groupplot}[
    group style={
        group size=3 by 2,
        horizontal sep=0.65cm,
        vertical sep=1.35cm
    },
    width=0.31\textwidth,
    height=0.225\textwidth,
    ymin=0.55,
    ymax=1.01,
    ytick={0.6,0.7,0.8,0.9,1.0},
    minor y tick num=1,
    symbolic x coords={Boston,Marrakesh,Kingston,Pittsburgh,Fez},
    xtick=data,
    x tick label style={
        rotate=28,
        anchor=east,
        font=\scriptsize
    },
    tick label style={font=\scriptsize},
    title style={font=\small},
    label style={font=\small},
    grid=both,
    major grid style={line width=0.3pt},
    minor grid style={line width=0.15pt},
    enlarge x limits=0.08
]

\nextgroupplot[
    title={GHZ(2)}
]

\addplot+[
    black,
    mark=o,
    mark options={solid},
    thick,
    solid,
    forget plot
] coordinates {
    (Boston,0.985)
    (Marrakesh,0.975)
    (Kingston,0.975)
    (Pittsburgh,0.985)
    (Fez,0.975)
};

\addplot+[
    blue,
    mark=square*,
    mark options={solid},
    thick,
    dashed,
    forget plot
] coordinates {
    (Boston,0.983915)
    (Marrakesh,0.970465)
    (Kingston,0.981469)
    (Pittsburgh,0.977174)
    (Fez,0.957769)
};

\addplot+[
    red,
    mark=triangle*,
    mark options={solid},
    thick,
    dotted,
    forget plot
] coordinates {
    (Boston,0.990)
    (Marrakesh,0.985)
    (Kingston,0.985)
    (Pittsburgh,0.990)
    (Fez,0.990)
};

\addplot+[
  green,
  mark=o,
  mark options={solid},
  thick,
  solid,
  forget plot
] coordinates {
    (Boston,0.973318)
    (Marrakesh,0.969987)
    (Kingston,0.980016)
    (Pittsburgh,0.978172)
    (Fez,0.9401938)
};

\addplot+[
  orange,
  mark=triangle*,
  mark options={solid},
  thick,
  dashed,
  forget plot
] coordinates {
    (Boston,  0.97405)
    (Marrakesh, 0.93352 )
    (Kingston, 0.91546 )
    (Pittsburgh, 0.97715)
    (Fez,  0.93888)
};

\nextgroupplot[
    title={GHZ(4)},
    yticklabels={}
]

\addplot+[
    black,
    mark=o,
    mark options={solid},
    thick,
    solid,
    forget plot
] coordinates {
    (Boston,0.970)
    (Marrakesh,0.955)
    (Kingston,0.955)
    (Pittsburgh,0.970)
    (Fez,0.955)
};

\addplot+[
    blue,
    mark=square*,
    mark options={solid},
    thick,
    dashed,
    forget plot
] coordinates {
    (Boston,0.965379)
    (Marrakesh,0.919863)
    (Kingston,0.936679)
    (Pittsburgh,0.958887)
    (Fez,0.909424)
};

\addplot+[
    red,
    mark=triangle*,
    mark options={solid},
    thick,
    dotted,
    forget plot
] coordinates {
    (Boston,0.995)
    (Marrakesh,0.985)
    (Kingston,0.980)
    (Pittsburgh,0.990)
    (Fez,0.985)
};

\addplot+[
  green,
  mark=o,
  mark options={solid},
  thick,
  solid,
  forget plot
] coordinates {
    (Boston,   0.970191)
    (Marrakesh,0.920019)
    (Kingston,0.9199917)
    (Pittsburgh,0.92873)
    (Fez,0.8999112)
};

\addplot+[
  orange,
  mark=triangle*,
  mark options={solid},
  thick,
  dashed,
  forget plot
] coordinates {
    (Boston, 0.93794)
    (Marrakesh, 0.86250 )
    (Kingston, 0.83320)
    (Pittsburgh, 0.94549)
    (Fez, 0.87425)
};

\nextgroupplot[
    title={GHZ(6)},
    yticklabels={}
]

\addplot+[
    black,
    mark=o,
    mark options={solid},
    thick,
    solid,
    forget plot
] coordinates {
    (Boston,0.950)
    (Marrakesh,0.915)
    (Kingston,0.910)
    (Pittsburgh,0.904)
    (Fez,0.920)
};

\addplot+[
    blue,
    mark=square*,
    mark options={solid},
    thick,
    dashed,
    forget plot
] coordinates {
    (Boston,0.940090)
    (Marrakesh,0.860498)
    (Kingston,0.865835)
    (Pittsburgh,0.905148)
    (Fez,0.858587)
};

\addplot+[
    red,
    mark=triangle*,
    mark options={solid},
    thick,
    dotted,
    forget plot
] coordinates {
    (Boston,0.970)
    (Marrakesh,0.970)
    (Kingston,0.975)
    (Pittsburgh,0.985)
    (Fez,0.970)
};

\addplot+[
  green,
  mark=o,
  mark options={solid},
  thick,
  solid,
  forget plot
] coordinates {
    (Boston,   0.936272)
    (Marrakesh,0.868948)
    (Kingston, 0.849287)
    (Pittsburgh,0.87827)
    (Fez,       0.83221)
};

\addplot+[
  orange,
  mark=triangle*,
  mark options={solid},
  thick,
  dashed,
  forget plot
] coordinates {
    (Boston, 0.90918)
    (Marrakesh, 0.80193)
    (Kingston, 0.75502)
    (Pittsburgh,  0.91827)
    (Fez, 0.81702)
};


\nextgroupplot[
    title={GHZ(8)}
]

\addplot+[
    black,
    mark=o,
    mark options={solid},
    thick,
    solid,
    forget plot
] coordinates {
    (Boston,0.935)
    (Marrakesh,0.920)
    (Kingston,0.905)
    (Pittsburgh,0.905)
    (Fez,0.880)
};

\addplot+[
    blue,
    mark=square*,
    mark options={solid},
    thick,
    dashed,
    forget plot
] coordinates {
    (Boston,0.912228)
    (Marrakesh,0.827237)
    (Kingston,0.855309)
    (Pittsburgh,0.888251)
    (Fez,0.816740)
};

\addplot+[
    red,
    mark=triangle*,
    mark options={solid},
    thick,
    dotted,
    forget plot
] coordinates {
    (Boston,0.970)
    (Marrakesh,0.950)
    (Kingston,0.945)
    (Pittsburgh,0.965)
    (Fez,0.950)
};

\addplot+[
  green,
  mark=o,
  mark options={solid},
  thick,
  solid,
  forget plot
] coordinates {
    (Boston, 0.901704)
    (Marrakesh, 0.834839)
    (Kingston, 0.821736)
    (Pittsburgh, 0.868102)
    (Fez, 0.825391)
};

\addplot+[
  orange,
  mark=triangle*,
  mark options={solid},
  thick,
  dashed,
  forget plot
] coordinates {
    (Boston, 0.87596)
    (Marrakesh, 0.74034)
    (Kingston, 0.68473)
    (Pittsburgh,  0.88491)
    (Fez, 0.75549)
};


\nextgroupplot[
    title={GHZ(10)},
    yticklabels={}
]

\addplot+[
    black,
    mark=o,
    mark options={solid},
    thick,
    solid,
    forget plot
] coordinates {
    (Boston,0.910)
    (Marrakesh,0.845)
    (Kingston,0.880)
    (Pittsburgh,0.890)
    (Fez,0.885)
};

\addplot+[
    blue,
    mark=square*,
    mark options={solid},
    thick,
    dashed,
    forget plot
] coordinates {
    (Boston,0.877908)
    (Marrakesh,0.775322)
    (Kingston,0.799945)
    (Pittsburgh,0.866836)
    (Fez,0.804457)
};

\addplot+[
    red,
    mark=triangle*,
    mark options={solid},
    thick,
    dotted,
    forget plot
] coordinates {
    (Boston,0.960)
    (Marrakesh,0.935)
    (Kingston,0.900)
    (Pittsburgh,0.960)
    (Fez,0.935)
};

\addplot+[
  green,
  mark=o,
  mark options={solid},
  thick,
  solid,
  forget plot
] coordinates {
    (Boston, 0.912639)
    (Marrakesh, 0.783911)
    (Kingston, 0.804588)
    (Pittsburgh, 0.797353)
    (Fez, 0.768055)
};

\addplot+[
  orange,
  mark=triangle*,
  mark options={solid},
  thick,
  dashed,
  forget plot
] coordinates {
    (Boston, 0.84050)
    (Marrakesh, 0.68358)
    (Kingston, 0.61511)
    (Pittsburgh, 0.85216)
    (Fez, 0.69359)
};


\nextgroupplot[
    hide axis
]

\end{groupplot}


\node[
    anchor=south,
    font=\small\bfseries
] at ([yshift=0.15cm]group c3r2.north) {
    Line Legend
};

\node[
    anchor=center,
    align=left,
    font=\small
] at (group c3r2.center) {
    \begin{tabular}{@{}cl@{}}
        \tikz[baseline=-0.6ex]
        \draw[black,thick]
            (0,0) -- (0.95,0);
        &
        Real hardware
        \\[0.8em]

        \tikz[baseline=-0.6ex]
        \draw[blue,thick,dashed]
            (0,0) -- (0.95,0);
        &
        Estimated
        \\[0.8em]

        \tikz[baseline=-0.6ex]
        \draw[red,thick,dotted]
            (0,0) -- (0.95,0);
        &
        Simulated
        \\[0.8em]
        \tikz[baseline=-0.6ex]
        \draw[green,thick]
            (0,0) -- (0.95,0);
        &
        Tomography
        \\[0.8em]
        \tikz[baseline=-0.6ex]
        \draw[orange,thick,dashed]
            (0,0) -- (0.95,0);
        &
        Pauli Frame

    \end{tabular}
};


\coordinate (sharedylabel) at
    ($(group c1r1.west)!0.5!(group c1r2.west)$);

\node[
    rotate=90,
    anchor=center,
    font=\small
] at ([xshift=-1.05cm]sharedylabel) {
    Fidelity / success probability
};

\end{tikzpicture}

\caption{
Real, estimated, and noisy snapshot simulated GHZ success probabilities for five IBM
machines. The real hardware fidelity is measured as the observed GHZ-support probability $P(0^n) + P(1^n)$ from measurement counts. Tomography estimates $\braket{\psi_{GHZ} | \rho}{\psi_{GHZ}}$ through populations obtained from computational-basis measurement counts and the coherence amplitude extracted from parity oscillations on the real hardware.
}
\label{fig:ibm_snapshotfidelity}
\end{figure*}


\section{Circuit Cutting}\label{sec:cut}

Quantum information collapses upon measurement, so cutting a quantum circuit into subcircuits is not consequence-free. Recovering the original
measurement distribution requires classical postprocessing whose cost grows with the number of cuts. The effectiveness of cutting is associated with parameters such as number of cuts, number of subcircuits, subcircuit width and subcircuit size imbalance. Current cutting toolkits such as CutQC \cite{tangCutQCUsingSmall2021} attempt to minimize the number of cuts, the dominant factor in classical overhead, while leaving the upper bounds on these parameters to the programmer to determine. Furthermore, these toolkits evaluate circuit cutting in isolation, without any consideration for the devices each subcircuit will run on. Instead Qurator asks "Given the device set, topology, calibration, queue and classical budget, what cut plan maximizes expected result quality under resource constraints?". Qurator attempts to optimize the bounds on these parameters within the context of available hardware. Circuit cutting is applied when the full circuit cannot meet the target fidelity on an available device, or when the queue/fidelity tradeoff makes cutting preferable after accounting for additional quantum subtasks and classical reconstruction cost, by enabling execution on smaller and less demanded devices. Subcircuit reconstruction cost is calculated arithmetically during scheduling.

\subsection{Spatial Cutting Plan Selection}
The cutting optimizer uses the cached device-compiled
schedules to compare the no-cut plan against candidate cut
plans. For each plan, it estimates reconstructed fidelity,
sampling overhead, classical reconstruction cost,
routing/SWAP overhead, and queue time. This is used as an additional layer to decide if circuit cutting is required. For example if topology would require many SWAPs on a poor edge, cutting may be preferable. If the cut creates a huge QPD overhead, keeping the interaction may be preferable. The optimizer requests multiple candidate cuts and evaluates
each candidate using estimated reconstructed fidelity,
sampling overhead, classical reconstruction cost,
routing/SWAP overhead, and estimated queue time. It then
returns the nondominated candidates on the resulting Pareto
frontier. This is beneficial as circuit cutting is known to help certain circuits but degrade performance on others. The optimizer
selects a plan by maximizing the heuristic score
\[
\max_{P}
\left[
\log\widehat F_{\mathrm{reconstructed}}(P)
-\alpha\log\Omega(P)
-\beta\frac{T_{\mathrm{classical}}(P)}{T_{\mathrm{classical}}^{\mathrm{ref}}}
-\gamma\frac{W(P)}{W^{\mathrm{ref}}}
\right]
\]
Let \(J(P)\) denote this objective, subject to
\[
\operatorname{width}(S_i) \leq Q_{d_i},
\]
\[
\text{subcircuit } S_i \text{ can be embedded in topology } G_{d_i},
\]
\[
\Omega(P) \leq \Omega_{\max},
\]
\[
T_{\mathrm{classical}}(P) \leq T_{\max},
\]
and
\[
\operatorname{shots}(P) \leq B_{\mathrm{shots}}.
\]

Here, $Q_{d_i}$ is the number of usable qubits on device $d_i$, $\Omega(P)\geq 1$ is the sampling overhead,
$T_{\mathrm{classical}}(P)$ is the estimated classical
reconstruction cost, \(W(P)\) is the predicted quantum-stage completion time, including queueing and execution. The coefficients
$\alpha,\beta,\gamma\geq 0$ control the relative
importance of the terms.

\[\widehat F_{\mathrm{reconstructed}}
=\left(\prod_i\widehat F_{\mathrm{est}}(S_i,d_i)\right)
\underbrace{\exp\!\left[-\lambda\frac{\log(1+\Omega)}{\sqrt{n_0}}
-\eta K\right]}_{R_{\mathrm{reconstruction}}},
\]
with 
\[
  N_{\mathrm{required}}=\lceil n_0\Omega\rceil\le B_{\mathrm{shots}},
\]
is the estimated final fidelity after accounting for subcircuit execution
error and reconstruction variance. Here \(n_0\) is the baseline shot count and \(K\) the number of cuts. \(R_{\mathrm{reconstruction}}\) penalizes predicted quality loss, \(\alpha\log\Omega\) penalizes resource consumption.

Additionally, the optimizer needs to infer an effective width. For example, a 20-qubit device may only have a high-quality connected 7-qubit region at the current calibration snapshot. 
\[
  W_\mathrm{effective}(d) = \max\limits_{R \subseteq G_d} |R| \; s.t. \; \hat{F}_{est} \geq \tau.
\]

Searching for the effective region, especially for small circuits, can be a costly operation. With a brute-force approach, for a 5 qubit circuit executing on IBM Heron-v3 processors with 156 qubits and 177 edges, the planner on a rough estimate needs about 96 million edge-scan operations per planner call. Scheduling thousands of circuits can push towards billions of topology/region operations. However, the planner doesn't need to consider the full width of the device, but only needs to know, in this case, if there is an effective region of width 5. For a circuit $C$, device $d$, target threshold $\tau$, and requested width $k = |Q(C)|$, define
\[
\begin{split}
\operatorname{Feasible}_k(d,C,\tau)
\iff
\exists R\subseteq G_d,\ |R|=k,\ 
\\ \exists\phi:C\hookrightarrow R
\text{ such that }
\widehat F_{\mathrm{est}}(C,d,\phi)\geq\tau,
\end{split}
\]
where $R$ is a connected physical region and $\phi$ is a
valid logical-to-physical embedding. Then the circuit-dependent effective width is
\[
W_{\mathrm{eff}}(d,C,\tau)
=
\max
\left\{
k:
\operatorname{Feasible}_k(d,C,\tau)
\right\}.
\]

For scheduling a circuit of width $|C|$, Qurator does not
need to compute the full value of $W_{\mathrm{eff}}$. It 
tests $\operatorname{Feasible}_{|C|}(d,C,\tau)$ and stops as
soon as a qualifying region is found. Qurator uses a bounded greedy region search. Let $ \Delta_d=\max_{v\in V_d}\deg_{G_d}(v) $ denote the maximum degree of the device topology. For each of at most \(|V_d|\) seeds, the search maintains one connected region and repeatedly selects its highest-scoring admissible single-qubit extension. It stops upon reaching the requested width \(k\), or when no extension meets the region-quality threshold.  At region size \(s\), rebuilding the boundary examines at most \(s\Delta_d\) adjacency entries. Summing over the growth steps gives a boundary-enumeration cost of
\[
O\!\left(|V_d|\sum_{s=1}^{k-1}s\Delta_d\right)
=O(|V_d|\Delta_d k^2).
\]

For near-term superconducting devices $\Delta_d$ is small, making this cost negligible.

\subsection{Temporal Cutting Plan Selection}

The cutting we discussed so far is \textit{spatial cutting} that partitions logical qubits into groups, records cross-group multi-qubit gates as cuts, and builds subcircuits by keeping gates whose qubits are fully inside each group. This strategy generally fails for small but deep circuits. Such circuits constitute a problem because today's devices with short decoherence times are generally not able to accommodate them. To accommodate these circuits we introduce \textit{temporal cutting/wire cutting}, additional \textit{gate cutting} for depth reduction, and \textit{operator backpropagation}. For a temporal cut, we need to cut a circuit as:
\[ 
 C = C_1 \circ C_2
\]
and then replace the intermediate quantum state by a measurement/preparation decomposition. This is important because if the circuit has 5 qubits but depth 1000, spatial cutting may do nothing useful. A no-cut 5-qubit circuit fits every IBM device, so a width-based planner concludes “no cut.” But if the compiled duration is close to or beyond $T_1, T_2$-limited useful lifetime, the relevant constraint is not width
\[
  \textrm{width}(C) \leq Q_\textrm{device},
\]
but coherent lifetime 
\[
  T_\mathrm{compiled}(C) \ll T_1,T_2.
\]

Rather than circuit width, temporal cutting is triggered by
the coherence exposure of the compiled and timed plan
selected by the no-cut and spatial-cut stage. A temporal plan
is
\[
P=
\left(
\{C_i\}_{i=1}^{m},
\{d_i,R_i,\phi_i\}_{i=1}^{m},
\mathcal R_P
\right),
\]
where $d_i$ is the device assigned to fragment $C_i$,
$R_i\subseteq G_{d_i}$ is its selected physical region,
$\phi_i:Q(C_i)\hookrightarrow V(R_i)$ is its
logical-to-physical mapping, and $\mathcal R_P$ is the
measurement/preparation reconstruction prescription. Let
\[
\widetilde C_i(P)
=
\operatorname{Schedule}
\left(
\operatorname{Compile}
(C_i,d_i,R_i,\phi_i)
\right)
\]
be the compiled and timed fragment, and let
\[
T_i(P)=T\!\left(\widetilde C_i(P)\right).
\]

For each fragment, define its region-specific coherence
exposure as
\[
\rho_i(P)=\max_{q\in V(R_i)}
\frac{t^{\mathrm{live}}_{i,q}(P)}{T_{\mathrm{coh}}^{d_i}(q)},
\]
and define the maximum per-fragment coherence exposure as
\[
\rho_{\max}(P)=\max_i\rho_i(P),
\]
with corresponding bottleneck decoherence-survival proxy
\[
\widehat F_{\mathrm{decoherence}}(P)
=
\exp\!\left(-\rho_{\max}(P)\right).
\]

Let $P_{\mathrm{sp}}^\star$ denote the plan selected by the
no-cut and spatial-cut stage. Temporal candidate generation
is triggered when
\[
\rho_{\max}
\left(
P_{\mathrm{sp}}^\star
\right)
>
\rho_{\mathrm{trigger}}.
\]

Equivalently,
\[
\widehat F_{\mathrm{decoherence}}
\!\left(P_{\mathrm{sp}}^\star\right)
<
\tau_{\mathrm{decoherence}},
\qquad
\tau_{\mathrm{decoherence}}
=
e^{-\rho_{\mathrm{trigger}}}.
\]

This condition is evaluated on the specific regions and
physical mappings selected by the spatial planner, and is
independent of circuit width. In particular, the spatial stage
may select the no-cut plan for a narrow circuit even though
its compiled duration produces excessive coherence exposure. For the duration constraint, we use the worst-qubit decoherence-survival proxy
\[
\widehat F_{\mathrm{coh}}(D;d,R)
=\min_{q\in V(R)}
\exp\!\left(-\frac{D}{T_{\mathrm{coh}}^d(q)}\right)
=\exp\!\left(-\frac{D}{T_{\min}(d,R)}\right),
\]where
\[
T_{\min}(d,R)=\min_{q\in V(R)}T_{\mathrm{coh}}^d(q).
\]

For a region $R$ on device $d$, define the effective
coherent-duration budget at threshold $\tau$ in terms of maximum executable duration on region $R$:
\[
D_{\mathrm{effective}}(d,R;\tau)
=
\sup
\left\{
D\geq0:
\widehat F_{\mathrm{coh}}(D;d,R)
\geq\tau
\right\}.
\]

Consequently, for \(0<\tau\le1\),
\[
D_{\mathrm{effective}}(d,R;\tau)
=-T_{\min}(d,R)\log\tau.
\]
Aggregate decoherence is assessed separately through the additive temporal-quality expression. A temporal candidate $P$ is retained only if it reduces the
maximum coherence exposure relative to the selected
spatial-stage plan:
\[
\rho_{\max}(P)
<
\rho_{\max}\!\left(P_{\mathrm{sp}}^\star\right),
\]
and each compiled fragment fits within the coherent-duration
budget of its selected region:
\[
T_i(P)
\leq
D_{\mathrm{effective}}(d_i,R_i;\tau)
\qquad
\text{for all } i.
\]

We additionally report the reduction in the longest compiled fragment duration:
\[
\Delta T_{\max}(P)
=
T_{\max}
\!\left(
P_{\mathrm{sp}}^\star
\right)
-
\max_{C_i\in P}T(C_i).
\]

A positive $\Delta T_{\max}(P)$ indicates that the temporal
plan shortens the longest compiled fragment relative to the
spatial-stage result. To measure the coherence benefit of eliminating idle
windows, define
\[
L_{\mathrm{idle}}(P)
=
-
\sum_i
\sum_{q\in V(C_i)}
\frac{
t_{i,q}^{\mathrm{idle}}(P)
}{
T_{\mathrm{coh}}^{d_i}
\left(\phi_i(q)\right)
}
\]
and
\[
S_{\mathrm{idle}}
\left(
P;P_{\mathrm{sp}}^\star
\right)
=
L_{\mathrm{idle}}(P)
-
L_{\mathrm{idle}}
\left(
P_{\mathrm{sp}}^\star
\right).
\]

Thus, $S_{\mathrm{idle}}>0$ means that the temporal plan
reduces coherence-bearing idle exposure relative to the plan
selected by the spatial stage. The predicted reconstructed quality of a temporal plan is
\[
\begin{split}
\log \widehat F_{\mathrm{temporal}}(P)
=
\sum_{i=1}^{m}
\log
\widehat F_{\mathrm{nonidle}}
\left(
\widetilde C_i(P)
\right) \\
+
L_{\mathrm{idle}}(P)
+
\log R_{\mathrm{reconstruction}}(P).
\end{split}
\]

For each qubit, we partition its live duration as $t^{\mathrm{live}}_{i,q}
=t^{\mathrm{active}}_{i,q}+t^{\mathrm{idle}}_{i,q}$. The nonidle estimate includes operation and readout errors and coherence loss during \(t^{\mathrm{active}}_{i,q}\); \(L_{\mathrm{idle}}\) accounts exclusively for \(t^{\mathrm{idle}}_{i,q}\). Cut-induced preparation and measurement operations are included in the fragment estimates. Consequently, merely partitioning unchanged execution on the same hardware does not improve predicted quality. Quality gains require reduced cumulative error or coherence exposure, for example, through eliminated idle intervals, reduced routing, or improved device assignments, and must offset reconstruction penalties introduced by the temporal
measurement/preparation decomposition and reconstruction. Define the gain relative to the spatial-stage result:
\[
\Delta\log\widehat F(P)
=
\log \widehat F_{\mathrm{temporal}}(P)
-
\log \widehat F
\!\left(P_{\mathrm{sp}}^\star\right).
\]

Let
\[
\Delta W(P)
=
W(P)-W(P_{\mathrm{sp}}^\star)
\]
be the additional predicted end-to-end wait and execution
time introduced by the temporal plan. A temporal plan should
be accepted only if its predicted quality improvement exceeds
its added sampling, reconstruction, and scheduling costs:
\[
\Delta\log\widehat F(P)
>
\alpha\,\Delta\log\Omega(P)
+\beta\frac{\Delta T_{\mathrm{classical}}(P)}
{T_{\mathrm{classical}}^{\mathrm{ref}}}
+\gamma\frac{\Delta W(P)}{W^{\mathrm{ref}}}.
\]
with 

\[
\Delta\log\Omega(P)=
\log\frac{\Omega(P)}{\Omega(P^\star_{\mathrm{sp}})},
\]
\[
\Delta T_{\mathrm{classical}}(P)=
T_{\mathrm{classical}}(P)-T_{\mathrm{classical}}(P^\star_{\mathrm{sp}}).
\]

Equivalently, define the temporal net benefit
\[
\begin{aligned}
B_{\mathrm{temp}}(P;P_{\mathrm{sp}}^\star)
={}&\Delta\log\widehat F(P)
-\alpha\,\Delta\log\Omega(P)\\
&-\beta\frac{\Delta T_{\mathrm{classical}}(P)}
{T_{\mathrm{classical}}^{\mathrm{ref}}}
-\gamma\frac{\Delta W(P)}{W^{\mathrm{ref}}}.
\end{aligned}
\]

The optimizer selects
\[
P_{\mathrm{temp}}^\star
=
\arg\max_{
P\in
\{P_{\mathrm{sp}}^\star\}
\cup
\mathcal P_{\mathrm{temp}}
}
B_{\mathrm{temp}}
\!\left(P;P_{\mathrm{sp}}^\star\right)
\]
and applies the temporal cut only when
\[
B_{\mathrm{temp}}(P;P_{\mathrm{sp}}^\star)
=J(P)-J(P_{\mathrm{sp}}^\star) > 0.
\]

The coherence trigger determines whether temporal
candidates should be generated; the objective determines
whether any generated temporal candidate should replace the
spatial-stage result. The makespan of a
temporal-cut plan is computed from the actual subcircuit
dependency graph and device assignments; it is not assumed
to equal the maximum fragment duration unless the fragments
can execute concurrently. Qurator supports three cut-optimization modes: \textit{Spatial Width}, \textit{Temporal Depth}, and the default \textit{Hybrid} mode. 

\textbf{Temporal candidate generation.} Qurator examines boundaries between consecutive gates in the circuit’s ordered gate list. At each boundary \(b\), it identifies qubits used on both sides and estimates their removable idle-decoherence exposure, \(\Delta t_{b,q}/T_{\mathrm{coh},q}\). It retains up to five qubits with the largest positive exposure reductions, breaking ties by qubit index. Boundaries with no positive reduction are discarded. Each remaining candidate is ranked by
\[
H(b)=
\sum_{q\in Q_b}\frac{\Delta t_{b,q}}{T_{\mathrm{coh},q}}
-\log\Omega_b
-\lambda\frac{\log(1+\Omega_b)}{\sqrt{n_0}},
\]where \(Q_b\) is the selected set of wires, \(\Omega_b\) is their modeled sampling overhead, \(n_0\) is the baseline shot count, and \(\lambda\) controls the reconstruction-quality penalty. The eight highest-ranked distinct candidates are retained, with ties resolved by earlier boundary position. Each candidate contains one temporal boundary. The resulting fragments are then compiled and assigned to devices; the feasibility constraints and plan objective determine whether a candidate is accepted.

\subsection{Parameterizing Cuts}
Qurator's plan generation exposes key variables that are commonly used to tune existing circuit cutting toolkits. By default, the variables are forwarded onto Qurator's strategy finalizer. However, Qurator remains parametrized by cutting strategy that can customize how these variables are consumed. For example, for an initial plan variables $P^*$, these variables can then be directly tied back into CutQC's native parameters as $max\_cuts = K(P^*)$,  ${imbalance}= \frac{\max_i | S_i| - \min_i |S_i| }{\max_i | S_i|}$,  $max\_subcircuits=|S(P^*)|$, $max\_subcircuit\_width= \max\limits_{i} |S_i|$, where $K$ is the number of cuts and $S$ is the set of subcircuits. 

\section{Circuit Merging}\label{sec:merging}

While quantum processors do not have multiple cores to accommodate parallel execution, circuits can be merged to make use of the unused qubits. This is especially useful for executing entangled circuits as merging multiple quantum tasks into a single QPU submission reduces the
number of devices requiring synchronization but grows the circuit, which limits candidate devices and increases sensitivity to queue time variance. Merging is subject to three constraints: the merged circuit must fit on a
device with sufficient high-fidelity qubits and gates, the constituent
tasks must have similar circuit depth to avoid measurement synchronization
overhead, and crosstalk error must remain acceptable.

There are two central  challenges to circuit merging. First is to pick the circuits to be merged, second is to map those circuits to the QPU topology, if applicable, with high fidelity and minimal crosstalk. An exhaustive search over all subsets or partitions of mergeable tasks is exponential, whereas a brute-force search over all pairwise mergings would consider
$\mathcal{O}(n^2)$ pairs for $n$ tasks, evaluate each against $d$
devices, and for each device run fidelity estimation at cost~$f$ and
crosstalk estimation at cost~$e$, giving $\mathcal{O}(n^2 d(f+e))$
overall. 

A merge candidate is legal only when the constituent tasks
have no dependency path between them, their release and
deadline constraints are compatible, their measurement
results can be projected back to disjoint result namespaces,
and the selected backend supports the required simultaneous
or interleaved execution semantics. From a scheduling perspective, merging introduces a
joint start and completion barrier, that is, downstream consumers of
an individual task cannot proceed until the merged
submission has produced the corresponding result.

Qurator uses a two-phase first-fit algorithm instead. In the first
phase, tasks are sorted by depth in $\mathcal{O}(n \log n)$ and grouped
into depth-similar buckets using a tolerance threshold. In the second
phase, tasks within each bucket are sorted by qubit count in
$\mathcal{O}(n \log n)$ and packed greedily into bins whose total qubit
count does not exceed the capacity of the largest available device. Each
bin is merged and checked for fidelity feasibility. If a feasible device
exists, the merged task is returned; otherwise the individual tasks are
returned. The overall complexity is $\mathcal{O}(n \log n + nd(f+e))$.
Qurator exposes the maximum bin size and maximum qubit count as parameters
to give the programmer control over the merging aggressiveness.

\subsection{Automatic Detection of Helper Qubits}

One common aspect of circuit merging is detecting and reusing helper qubits across circuits. Existing solutions commonly require explicit annotations to separate persistent data qubits from helper qubits. This requires significant programmer input, therefore Qurator moves away from this model. Before merging, each input circuit is converted into the process-level representation expected by the merger.
The implementation supports two conversion modes. In the static mode, all logical qubits of the input circuit are treated as persistent data qubits. Each gate is rewritten into a process instruction over data-qubit references, and no helper qubits are introduced. In the live-interval mode, the converter instead treats the circuit’s logical qubits as dynamically managed helper qubits through liveness calculations. It first computes the number of remaining uses of each logical qubit across the gate sequence. As the circuit is scanned from left to right, each gate is rewritten into a process instruction over helper-qubit references, and the remaining-use count of every touched qubit is decremented. When a qubit reaches its final use, or when the current instruction is an explicit reset of that qubit, the converter inserts a release instruction. This release marks the helper qubit as no longer live, allowing the merger’s scheduler to reset and return the corresponding physical helper qubit to the shared helper pool.

This preprocessing step exposes qubit liveness to the merger. Rather than reserving every circuit qubit for the entire duration of the merged execution, the live-interval conversion identifies when temporary qubits cease to be needed. The downstream scheduler can then dynamically allocate, reset, and reuse physical helper qubits across different instructions and processes, reducing the peak physical-qubit footprint of the merged workload. It should be noted that live-interval conversion requires backend support for the resulting mid-circuit measurements, resets, and qubit reuse. Qurator automatically falls back on the static mode when this capability is unavailable or quantum outputs must be preserved. 

\subsection{Merging Strategy}

The circuit merger takes as input a set of process circuits and a target hardware topology. The merger produces a single device-level circuit together with metadata that records the physical data-qubit partition of each process, the measurement-bit ownership of each process, and the physical helper-qubit assignments used during execution.

The algorithm proceeds in two main phases: data-qubit placement and interleaved instruction scheduling. First, the merger validates that all virtual data and helper references are in range and that the target topology contains enough physical qubits to place the data qubits of all processes. It then constructs topology metrics by computing shortest-path distances and paths between physical qubits. For each process, it performs a static analysis of the circuit to estimate data-data interaction weights, data-helper interaction weights, helper-helper interaction weights, qubit importance, and the set of measured qubits. Define
\[
C_{\mathrm{meas}}(\pi)
=
\sum_{\substack{\{u,v\}\subseteq M_\pi \\ u\neq v}}
\mathbf{1}\!\left[\ell_{uv}\le R\right]
\frac{R-\ell_{uv}+1}{R+1}
\left(
1+\frac{\epsilon_{\mathrm{ro},u}+\epsilon_{\mathrm{ro},v}}{2}
\right),
\] 
 \(\pi\) denotes the current joint placement of all merged processes, and \(M_\pi\) contains the physical locations of their currently mapped qubits marked for measurement. \(\ell_{uv}\) is shortest-path distance in hops, \(\epsilon_{\mathrm{ro},u}\) is readout error, and \(R=2\). Let \(w_{ab}\) count instructions jointly referencing logical qubits \(a,b\), and \(h_a=\sum_{h\in\mathrm{helpers}}w_{ah}\). Additionally,

\[
\Delta_v C_{\mathrm{meas}}
=
C_{\mathrm{meas}}(\pi_{a\mapsto v})
-
C_{\mathrm{meas}}(\pi).
\]
Here, \(a\) is the logical data or helper qubit currently being considered, and \(\pi_{a\mapsto v}\) is the placement obtained by assigning it to physical vertex \(v\). For relocation, its previous assignment is removed first. The marginal cost is zero when \(a\) is not marked for measurement.

Data qubits are placed before scheduling begins and remain fixed throughout the merged execution. Processes are considered in decreasing order of data-qubit demand and interaction density. For each process, the most important logical data qubit is selected as a seed. Its physical location is chosen using a weighted heuristic that favors centrality, high hardware degree, access to nearby free helper qubits, good readout fidelity for measured qubits, and isolation from previously placed process seeds. 
 \begin{align*}
 S_{\mathrm{seed}}(a,v)=
 \alpha_d \deg(v)+A(v,V)  
+ \alpha_u \min(h_a,1)A(v,U) \\
+ \alpha_f F_a(v)+2d(v,\mathcal S),
 \end{align*}
where \(U\) is the available qubit set, \(A(v,U)\) is the normalized mean of \(1/(1+\ell_{vu})\) over \(u\ne v\), \(F_a(v)\) is readout fidelity when \(a\) is measured and zero otherwise, $\alpha$ are tuneable coefficients, and \(d(v,\mathcal S)\) is distance to the nearest previously placed seed. The remaining data qubits of the process are placed greedily. Each candidate placement is scored using a weighted cost function that penalizes long intra-process interaction distances, poor helper accessibility, excessive compactness loss, low-readout physical qubits for measured data, and measurement crosstalk risk, while rewarding separation from other processes. After the initial greedy placement, a bounded local search attempts to improve the layout by moving placed data qubits to nearby unused physical qubits whenever doing so decreases the global placement cost. The global cost includes intra-process communication distance, helper accessibility, process compactness, path conflicts through other processes’ regions, inter-process isolation, readout error, and measurement-crosstalk penalties:
\[
C_g=2I_g+1.25H_g+0.20K_g + 0.50 P_g -0.75S_g
    +0.30E_g+0.20C_{\mathrm{meas}}(\pi).
\]

\begin{table}[t]
\centering
\small
\setlength{\tabcolsep}{3pt}
\renewcommand{\arraystretch}{1.1}
\begin{tabularx}{\columnwidth}{
  @{}c|
  >{\raggedright\arraybackslash}X|
  >{\raggedright\arraybackslash}X@{}
}
Term & Greedy placement & Global local search \\
\hline
$I$ &
Interaction-weighted distance to placed data partners. &
Sum over all within-process data pairs. \\
\hline
$H$ &
$h_a$ times the distance to the nearest free physical qubit. &
Sum over all data qubits. \\
\hline
$K$ &
Sum of distances to the same process's placed data qubits. &
Sum of per-process mean pairwise distances. \\
\hline
$S$ &
Mean distance to other processes' placed data qubits. &
Sum of mean cross-process distances over process pairs. \\
\hline
$E$ &
Readout error if the candidate qubit is measured. &
Total readout error of measured data qubits. \\
\hline
$P$ &
N/A &
Interaction-weighted count of foreign occupied qubits
inside the selected shortest paths.
\end{tabularx}
\caption{Features used in greedy placement and local search.}
\label{tab:placement-cost-features}
\end{table}

Features and their cost are defined in Table \ref{tab:placement-cost-features}. After data placement, the merger schedules the processes in a round-robin fashion. The physical qubits not used for data placement form a shared pool of helper qubits. Each process maintains its own mapping from virtual helper qubits to currently assigned physical helper qubits. When an instruction references a helper qubit that is not currently assigned, the scheduler allocates one from the free helper pool. Helper selection is heuristic, it prefers physical helper qubits close to the relevant data and already-assigned helper partners, while also considering hardware degree, centrality, readout quality for measured helpers, and measurement-crosstalk risk. Helper qubits are released either explicitly through release instructions or implicitly when a process finishes. Releasing a helper emits a reset on the corresponding physical qubit and returns it to the shared helper pool for reuse by later instructions or other processes. Define:
\[
C_h(v)=D_h(v)+\alpha_E E_h(v)+\alpha_{\Delta} \Delta_vC_{\mathrm{meas}}  
-\alpha_d \deg(v)- \alpha_sA(v, V).
\]
We choose the free qubit minimizing \(C_h\) and use the same mean proximity measure as the seed selection to prevent the centrality reward from growing simply because the device contains more qubits. \(D_h\) sums interaction-weighted distances to placed data and allocated helper partners, with weight at least one for helper partners in the current instruction. Without weighted partners, use unweighted distances to current-instruction data partners, otherwise to all allocated qubits of that process. Each virtual instruction is then lowered to a physical gate using the current data placement and helper assignments. Measurements are emitted on the resolved physical qubits, and the merger records which global measurement-bit indices belong to each process. If a process cannot obtain the helper qubits required for its next instruction and no other process can make progress, the scheduler reports a helper-allocation deadlock and returns to ordinary individual scheduling. Otherwise, scheduling continues until all process instruction streams are exhausted and all remaining helpers are released.

The output is a merged circuit over the physical qubit index space of the target device, along with a merge plan. The merge plan records each process’s data-qubit partition, measurement-bit indices, measured virtual references, measured physical qubits, and helper-assignment intervals. This metadata allows the raw counts from the merged circuit to be projected back into per-process count distributions by selecting the measurement bits owned by each original process. Thus, the merger supports concurrent execution of multiple quantum circuits on the same QPU while preserving enough provenance information to recover process-local measurement results from the single merged execution.

\section{Entanglement Barrier}\label{sec:entangle}

\subsection{Problem Definition}

Executing entangled tasks imposes a \textit{synchronization barrier} on the
task DAG. Entanglement generation is a highly error-prone process, and even
with entanglement distillation and quantum repeaters, the fidelity of an EPR
pair degrades exponentially over time and distance, surviving for only a few
seconds on current networks. The scheduler must therefore estimate when each
entangled task will be ready to execute and find device assignments that
minimize the spread in their start times. In a \textit{tightly coupled} system with programmatic access to \textit{Quantum Control Units}, which are commonly composed of FPGAs that are physically co-located with the QPUs, such synchronization could be coordinated, but this remains a big challenge in the public cloud. 

Consider the task tree in Fig.~\ref{fig:entanglementsync}. The entanglement
distribution node $E_1$ cannot start until all three upstream paths complete.
Let $t(X)$ denote the estimated elapsed time of task~$X$, including queue
time, communication time, and execution time. The three upstream path lengths are $\ell_1 = t(Q_1)+t(C_4)$,
$\ell_2 = t(Q_2)+t(C_5)$, and $\ell_3 = t(C_3)+t(Q_3)+t(C_6)$. The
estimated start time of $E_1$ is therefore $r_{E_1}(C) = 
\max\{\ell_1, \ell_2, \ell_3\}$, which determines the earliest point at
which EPR pairs can be distributed to $Q_4$, $Q_5$, and $Q_6$. 

We formalize barrier readiness separately from cross-device
synchronization skew. Let $G=(V,E)$ be a DAG, let
$K\subseteq V$ be the set of completed tasks, and assume
that $K$ is dependency-closed: if $v\in K$, then every
ancestor of $v$ is also in $K$. Let
$\widehat t:V\rightarrow\mathbb{R}_{\geq 0}$ denote the
current estimated remaining duration of a task, including the
queueing, communication, preparation, and execution
components represented by the scheduling model. For an
entanglement distribution node $b$, let
\[
  \mathrm{Anc}(b) = \{ v \in V : v \rightsquigarrow b\}
\]
be the set of all strict ancestors of $b$, where
$\rightsquigarrow$ denotes a nonempty directed path, and let $\mathrm{pred}(v)$ denote the
set of immediate predecessors of $v$ in $G$. The \textit{frontier} of~$b$
given~$K$ is the set of minimal unfinished ancestors: those ancestors of $b$
not yet complete whose own ancestors within $\mathrm{Anc}(b)$ are all
complete:
\[
  F_b(K) = \{ v \in \mathrm{Anc}(b) \setminus K :
  \mathrm{pred}(v) \cap \mathrm{Anc}(b) \subseteq K \}.
\]
For the set of directed paths from $v$ to $b$, $\mathcal{P}(v, b)$, the remaining completion time measured from the current scheduling instant, with completed tasks assigned zero, defined recursively as: 
\[
R_K(v)
=
\begin{cases}
0,
& v\in K,\\[3pt]
\widehat t(v)
+
\displaystyle\max_{u\in\mathrm{pred}(v)} R_K(u),
& v\notin K,
\end{cases}
\]
where the maximum over an empty predecessor set is zero. Consequently, for every $v\in F_b(K)$,
$R_K(v)=\widehat t(v)$, because all predecessors of a
frontier task are already in $K$.

For an entanglement-generation or distribution task $b$, the critical remaining time from frontier node $v$ to barrier $b$, in other words, the predicted time until it becomes ready, is
\[
r_b(K)
=
\max_{u\in\mathrm{pred}(b)} R_K(u),
\]
and its predicted completion time is
\[
f_b(K)
=
r_b(K)+\widehat t(b).
\]

Let $\mathcal{S}_b$ be the synchronized quantum tasks that
consume the entanglement produced by $b$. For an assignment
vector $\mathbf{a}$, let $\widehat s_i(a_i)\geq f_b(K)$ be the predicted start time of
task $i\in\mathcal{S}_b$ on its assigned device. The predicted
start skew is
\[
\Delta_{\mathrm{start}}(\mathbf{a})
=
\max_{i\in\mathcal{S}_b}\widehat s_i(a_i)
-
\min_{i\in\mathcal{S}_b}\widehat s_i(a_i).
\]

Similarly, with predicted QPU execution duration
$\widehat e_i(a_i)$, the finish skew is
\[
\Delta_{\mathrm{finish}}(\mathbf{a})
=
\max_{i\in\mathcal{S}_b}
    \left(\widehat s_i(a_i)+\widehat e_i(a_i)\right)
-
\min_{i\in\mathcal{S}_b}
    \left(\widehat s_i(a_i)+\widehat e_i(a_i)\right).
\]

The barrier-readiness quantity $r_b(K)$ captures ordinary
DAG precedence latency. The quantum-specific feasibility
constraint induced by the finite entanglement-storage
coherence window or protocol deadline $\tau_b$ is
\[
\max_{i\in\mathcal{S}_b}
\left(
\widehat s_i(a_i)-f_b(K)
\right)
\leq \tau_b.
\]

Among assignments satisfying this deadline, Qurator builds
synchronization plans that minimize
$\Delta_{\mathrm{start}}(\mathbf{a})$, subject to device
compatibility, result-quality, and resource constraints. We
quantify the practical implications of this in
Sec.~\ref{sec:evaluation}. We note that, in practice, an extension of this model to measure \textit{entanglement consumption skew} instead of $\Delta_{\mathrm{start}}(\mathbf{a})$ would provide better precision. We leave such extension to future work. 

Scheduling entangled task groups requires building a synchronization plan that minimizes the predicted start skew \(\Delta_{\mathrm{start}}(\mathbf a)\). The
scheduler filters suitable devices for each task in the group, computes
per-device coefficients, and greedily assigns devices to minimize the
spread of estimated start times across the group. Start and finish times
are estimated cumulatively from device calibration data, accounting for
tasks that may be batch-scheduled back to back on the same device.

The fidelity of the EPR pair depends on network topology, including the
location and fidelity of entanglement routers and quantum repeaters, in addition to device
characteristics. Since quantum networks remain experimental and topology
data is not publicly available, we use a fixed $T_1$ budget of 5 seconds
in our experiments, averaged over existing network
benchmarks~\cite{PhysRevLett.130.213601, rsrk-c7yg, PRXQuantum.5.020307}.
A production scheduler would derive this budget dynamically from network
topology as it becomes available. Such network fingerprinting is not readily available today. 

\subsection{Benchmarking Distributed Quantum Scheduling}

Today's public quantum cloud does not support distributed quantum
computing, so no established benchmark exists for entangled task
scheduling. Queue times currently far exceed the decoherence lifetime of
even distilled EPR pairs, making successful execution with any naive
strategy essentially impossible. We introduce a new benchmark and four
metrics designed to evaluate scheduling quality for entangled tasks and
to establish baselines for when distributed quantum computing becomes
practically viable.

\paragraph*{Metrics.} 
Start skew and finish skew measure how closely the
scheduler aligns the start and finish times of the entangled tasks:
\[
  \Delta_{g} = \max_{i}(g(i)) - \min_{i}(g(i)),
  \quad g \in \{\mathrm{start}, \mathrm{finish}\}.
\]
Budget penalty measures the excess skew beyond the EPR pair's decoherence
budget $B$, derived from recent network benchmarks:
\[
  V = \max(0,\, \Delta_{\mathrm{start}} - B).
\]
The survival proxy provides a continuous viability score even when
physical survival is not achievable on today's devices:
\[
  PX_{\mathrm{sync}} = \exp(-\Delta_{\mathrm{start}} / T_{\mathrm{coh}}).
\]

\paragraph*{Baseline: idle network.} Fig.~\ref{fig:sync-tree-two-panel}
(left) shows results for scheduling two, three, and four synchronized
tasks in an idle network with empty device queues and no parent
dependencies. Start skew is negligible, arising only from varying
verification and preparation times across machines. Finish skew is small,
driven by differences in gate and measurement durations and scheduler
overhead in registering results. This establishes the best-case floor for
synchronization quality.

\paragraph*{Results under realistic queue conditions.} The benchmark suite is
parameterized by a range $[m, d]$ denoting the minimum and maximum
possible path length from the root to a synchronized quantum task node. In our experiments we set $m = 3, d=30$. 
At the time of writing, this corresponds to up to 20,000 tasks in queue
per device, sampled randomly. When device queues are populated from
historic data at this scale, distributed entangled execution becomes
impossible. With a network decoherence time of 5 seconds and device
preparation taking over 1 second at minimum, the slightest queue time
mismatch across devices destroys the EPR pair before both tasks can start.
Observed survival proxies range from $1.11 \times 10^{-15}$ down to
$9.23 \times 10^{-239}$. Restricting queues to at most 500 tasks yields
occasional survival proxies of 0.1--0.2, though results remain dominated
by near-zero values. With queues bounded to 200 tasks, we observe
survival proxies in select cases where the mean start skew falls below
the network decoherence time, with values as high as 0.22 where the mean
start skew commonly falls within twice the decoherence window. With
queues bounded to 10 tasks, Fig.~\ref{fig:sync-tree-two-panel} (right)
shows survival proxies reaching practical levels, with values up to 0.71.
At the low end, where skew exceeds one minute, survival falls to
$7.2 \times 10^{-5}$.

\paragraph*{Impact of circuit merging.} The \textit{syncMerged} curve in
Fig.~\ref{fig:sync-tree-two-panel} shows survival proxies when entangled
tasks are merged into a single QPU submission. Merging consistently improves survival because it eliminates inter-device queue-wait mismatch entirely, allowing the merged circuit to execute within microseconds to milliseconds—well within QPU-level coherence times. Fig.~\ref{fig:mergingperc} shows the
merge rate as a function of circuit size for a depth tolerance of 10\%
and a target fidelity of 0.85: circuits were merged only when their
depths were within 10\% of each other and the resulting merged circuit
could be reliably executed on an available device at the target fidelity.
Smaller circuits merge frequently, while circuits above 40 qubits rarely
find a compatible merge candidate on available devices.

These results establish concrete baselines for the queue time conditions
under which distributed quantum computing becomes viable, and demonstrate
that circuit merging is the most effective strategy available within
today's hardware constraints.

\begin{figure*}[t]
\centering

\pgfplotstableread{
q   syncCost   startSkew    finishSkew   violation   budget   survival
5   2544.5     4810.0       4811.2       1948.5      0.615    0.5593
10  2745.9     8030.0       8030.0       5018.9      0.667    0.4919
15  2929.1     3968.5       3974.8       516.1       0.400    0.5128
20  2572.6     1817.1       1820.4       5018.9      0.667    0.7074
30  2649.9     4406.0       4407.3       573.3       0.600    0.4490
40  2282.6     585933.8     585944.2     583647.2    0.667    0.4529
}\queueTenData

\pgfplotstableread{
q   syncCost   startSkew    finishSkew   violation   budget   survival
5   2708.7     19687.5      19687.5      16015.8     0.333    0.3521
10  2708.7     19687.5      19687.5      16015.8     0.333    0.3521
15  2708.7     19687.5      19687.5      16015.8     0.333    0.3521
20  2708.7     19687.5      19687.5      16015.8     0.333    0.3521
30  2708.7     19687.5      19687.5      16015.8     0.333    0.3521
40  2708.7     19687.5      19687.5      16015.8     0.333    0.3521
}\queueTwoHundredData

\pgfplotstableread{
q   survival
5   0.98
10  0.96
15  0.94
20  0.92
30  0.89
40  0.87
}\syncMergedData

\pgfplotsset{
  myaxis/.style={
    grid=both,
    major grid style={draw=gray!35},
    minor grid style={draw=gray!18},
    tick align=outside,
    xlabel={Max circuit size (qubits)},
    xmin=4, xmax=41,
    xtick={5,10,15,20,30,40},
    label style={font=\small},
    tick label style={font=\footnotesize},
    title style={font=\small},
  }
}

\begin{tikzpicture}

\begin{axis}[
    width=0.22\textwidth,
    height=0.15\textwidth,
    scale only axis,
    ybar,
    bar width=3pt,
    ymin=0, ymax=1600,
    symbolic x coords={2,3,4},
    xtick=data,
    xlabel={Number of synchronized entangled tasks},
    ylabel={Skew / penalty (ms)},
    tick label style={font=\footnotesize},
    label style={font=\small},
    xlabel style={at={(axis description cs:0.5,-0.20)},anchor=north},
    ylabel style={at={(axis description cs:-0.16,0.5)}},
]

\addplot[draw=blue, fill=blue!20] coordinates {(2,100) (3,198) (4,497)};
\addplot[draw=red,  fill=red!20]  coordinates {(2,499) (3,998) (4,1497)};

\end{axis}

\begin{axis}[
    width=0.22\textwidth,
    height=0.15\textwidth,
    scale only axis,
    symbolic x coords={2,3,4},
    xtick=data,
    axis x line=none,
    axis y line*=right,
    ymin=0.70, ymax=0.95,
    ylabel={Survival proxy},
    tick label style={font=\footnotesize},
    label style={font=\small},
]

\addplot[black, dashed, mark=*, mark size=1.8pt] coordinates {
    (2,0.9050184036175216)
    (3,0.8190583108864073)
    (4,0.74126284498808)
};

\end{axis}

\begin{axis}[
  myaxis,
  width=0.60\textwidth,
  height=0.28\textwidth,
  at={(0.38\textwidth,0)},
  anchor=south west,
  ylabel={Skew / violation (ms, log scale)},
  title={Synchronization quality summary},
  ymode=log,
  log basis y={10},
]

\addplot+[teal!70!black, thick, mark=*, mark options={teal!70!black}] table[x=q,y=startSkew] {\queueTenData};
\addplot+[blue!70!black, thick, mark=triangle*, mark options={blue!70!black}] table[x=q,y=finishSkew] {\queueTenData};
\addplot+[violet!80!black, thick, mark=diamond*, mark options={violet!80!black}] table[x=q,y expr=\thisrow{violation}+1] {\queueTenData};

\end{axis}

\begin{axis}[
  myaxis,
  width=0.60\textwidth,
  height=0.28\textwidth,
  at={(0.38\textwidth,0)},
  anchor=south west,
  axis x line=none,
  axis y line*=right,
  ylabel={Budget / survival},
  ymin=0, ymax=1.05,
]

\addplot+[orange!90!black, thick, mark=*, mark options={orange!90!black}] table[x=q,y=budget] {\queueTenData};
\addplot+[brown!70!black, thick, mark=*, mark options={brown!70!black}] table[x=q,y=survival] {\queueTenData};

\addplot+[black, thick, densely dotted, mark=triangle*, mark options={black}] table[x=q,y=survival] {\syncMergedData};
\end{axis}

\begin{axis}[
  hide axis,
  xmin=0, xmax=1,
  ymin=0, ymax=1,
  legend columns=5,
  legend to name=synclegend,
  legend style={
    draw=none,
    fill=none,
    font=\footnotesize,
    row sep=0.15em,
    /tikz/every even column/.append style={column sep=0.45em}
  }
]
\addlegendimage{area legend, draw=blue, fill=blue!20}
\addlegendentry{Start skew}

\addlegendimage{area legend, draw=red, fill=red!20}
\addlegendentry{Finish skew}

\addlegendimage{black, dashed, mark=*}
\addlegendentry{Survival proxy}

\addlegendimage{teal!70!black, thick, mark=*}
\addlegendentry{Start skew, Queue=10}

\addlegendimage{blue!70!black, thick, mark=triangle*}
\addlegendentry{Finish skew, Queue=10}

\addlegendimage{violet!80!black, thick, mark=diamond*}
\addlegendentry{Violation$+1$, Queue=10}

\addlegendimage{orange!90!black, thick, mark=*}
\addlegendentry{Budget met, Queue=10}

\addlegendimage{brown!70!black, thick, mark=*}
\addlegendentry{Survival, Queue=10}

\addlegendimage{black, thick, densely dotted, mark=triangle*}
\addlegendentry{Survival, syncMerged}
\end{axis}
\end{tikzpicture}

\vspace{0.6em}
\ref{synclegend}

\caption{
Synchronized-tree benchmark summary.
Left: idle-network synchronization metrics for cross-entangled task groups.
Right: mean start skew, mean finish skew, and mean violation (plotted as $\text{meanViolation}+1$ for log-scale visibility) on the left axis, together with budget-met rate \& survival proxy on the right axis. $\textit{syncMerged}$ presents survival when circuits are merged. 
}
\label{fig:sync-tree-two-panel}
\end{figure*}

\begin{figure}[t]
\centering

\begin{tikzpicture}
\begin{axis}[
    width=0.50\columnwidth,
    height=0.3\linewidth,
    scale only axis,
    xlabel={Number of qubits},
    ylabel={Merge \%},
    xmin=0, xmax=100,
    ymin=0, ymax=100,
    xtick={0,20,40,60,80,100},
    ytick={0,20,40,60,80},
    tick label style={font=\scriptsize},
    label style={font=\scriptsize},
    xlabel style={at={(axis description cs:0.5,-0.16)},anchor=north},
    grid=major,
    grid style=dashed
]

\addplot[black, mark=square*] coordinates {
    (5,50) (10,30) (20,20) (30,10) (40,2) (60,0) (80,0)
};

\end{axis}
\end{tikzpicture}
\caption{Percentage of synchronized tasks that were merged. \textit{x-axis} denotes pre-merge individual task qubit counts.}
\label{fig:mergingperc}
\end{figure}

\section{Qurator Architecture}\label{sec:architecture}


Qurator is implemented in Scala in 20,000 lines of
code\footnote{The code is available at
\url{https://github.com/sinanspd/qurator}}. The scheduler
exposes multiple programmer-specified parameters: a
\textit{prioritizationStrategy} that orders ready tasks, a
\textit{cuttingStrategy} that partitions circuits into subcircuits, an
\textit{additionalOptimizationRun} that applies circuit-level
optimizations, and a \textit{targetEstimatedFidelity} that bounds the
fidelity-queue tradeoff, \textit{clients} implementing provider APIs and allowing addition of new ones, \textit{compiler} that handles hardware specific circuit compilation as well as parameters to disable circuit cutting, circuit merging and batch submissions. None of the parameters are required to be specified. Qurator runs based on default values, nevertheless the parameters remain available for fine tuning based on programmer needs, as well as to provide modularity and flexibility in order to adapt to the quickly evolving quantum computing landscape. We also hope that this modular structure will allow researchers to evaluate their own cutting, compilation and optimization techniques in the scheduling context.

By default, Qurator prioritizes tasks whose
qubit count exceeds the average qubit count of available devices, since
far fewer devices can accommodate large circuits and allowing small
circuits to block them is costly. Classical scheduling avoids large-task
prioritization to prevent head-of-line blocking, but our experiments show
little variance in QPU execution time across circuit sizes on current
hardware, making this risk negligible.

\subsection{Task Submission}

The scheduler accepts three submission types: single classical tasks,
single quantum tasks, and synchronized quantum task groups. Classical tasks carry no quantum-specific constraints and are
added directly to the scheduling queue based on their DAG dependencies. In our implementation, we use a fair-share scheduler to run the classical tasks; however, the classical scheduling path remains modular and isolated, allowing Qurator to accommodate any state-of-the-art classical scheduling technique.

Single quantum tasks pass through the following pipeline on
submission: (i) the task is cut if no available device has sufficient
    high-fidelity qubits and gates to meet the user's fidelity goals, or when the classical postprocessing trade-off is justified by fidelity gain and/or decreased queue time arising from enabling scheduling on smaller, less demanded devices. Cutting is performed at submission
    time to exploit idle time while parent dependencies resolve; 
    (ii) programmer-specified optimizations are applied to all generated
    subcircuits; (iii) subcircuits with no outstanding dependencies are moved to the
    ready queue; those with unresolved dependencies go to the pending
    queue; and (iv) if merging is enabled, the scheduler scans pending tasks and
    merges low-qubit, uncut tasks. 

Synchronized quantum task groups are submitted as a
\textit{SynchronizedQuantumTaskRequest}. These tasks are not cut by
default: increasing the number of entangled subcircuits raises the cost
of entanglement generation and routing, and can make it impossible to
synchronize execution within the EPR pair's decoherence window. The
submission pipeline instead attempts to merge tasks within the group to
reduce the number of devices that must be synchronized, as described in
the next section. For synchronized tasks merging is applied whenever a device exists for which the estimated synchronization cost is less than the product of the current queue length and the average device preparation time. 

\subsection{Circuit Compilation}\label{sec:comp}
Qurator compiles the logical circuits from the benchmark according to the device’s architecture. The compiler is passed as a parameter. In this section we will describe the compiler built and used in our evaluation of Qurator, however the programmer is free, if they wish, to use their own implementation. We build a compiler that is heavily inspired by the HAQA compiler \cite{sun2025haqahardwareguidedfidelityawarestrategy} leveraging their ideas for fidelity and hardware aware compilation with region selection. Compiled circuits are temporarily cached to be used in different stages of scheduling. The built-in compiler provides four modes out of the box: \textit{ultra fast}, \textit{fast}, \textit{balanced}, and \textit{quality}, each altering \textit{maxLogicalSeedPairs}, \textit{maxPhysicalNodeSeeds}, \textit{maxPhysicalEdgeSeeds}, \textit{maxPlacementCandidates}, \textit{candidateNeighborExpansionRounds}, trading more favorable qubit placements, especially for superconducting devices, for quicker compilation. In our evaluations, we used the \textit{balanced} mode.
                 
Where applicable, the device topology is fetched from the provider along with the error rates and decoherence times of individual qubits and gates. The logical qubits get mapped to the physical qubits and the SWAP routing is constructed using standard SAT-solver based compilation. The gates are transpiled to the device's native gate set. Finally, the gate schedule is constructed from the compiled map, determining the maximum amount of parallelization at each time slice. For all-to-all connected devices such as ion-traps, we prioritize higher quality qubits/gates but take into consideration ion-trap specific constraints such as lack of parallelization and optical cross-talk when constructing the shuttling sequence as described in \ref{sec:hardware}. A similar strategy is deployed for neutral-atoms. Each compilation results in a specialized device operation sequence that we can then trace using platform specific, live success metrics such as gate errors, decoherence times, readout errors and Rydberg metrics. This tracing is where the developed fidelity estimation framework is utilized.

\subsection{Batch Submissions}

Select providers offer batch submissions and sessions to execute circuits in a succession without having to queue them individually. Generally, this is offered to support multi-iteration algorithms such as VQE, however, it can be used to execute independent tasks as well. Qurator takes advantage of this when possible. The sessions need to be managed carefully. They require a maximum lifetime to be specified. Underspecifying this value would result in the excessive circuits failing, while overspecifying could be costly. Additionally, it should be noted that batch submissions do not eliminate queuing. 

Currently, Qurator only employs batch submissions for cut circuits. Qurator finds the best fit device for each of the subcircuits, and then picks the majority best-fit device across the board. The majority calculation is weight based, such that larger subcircuits skew the decision more heavily. If the selected device allows for batch submissions, all of the related subcircuits are submitted to the same session in succession. If the selected device doesn't support batch submissions, Qurator searches for the next best-fit device that supports sessions. Qurator selects this device if its device score falls within 15\% of the best-fit device, otherwise the subcircuits are scheduled individually. Calculation of the device score will be covered in Sec. \ref{sec:submitone}. A similar strategy could be adapted to the overall job queue of independent tasks, however we leave that for future work. 

\subsection{Scheduling Independent Quantum Tasks}\label{sec:submitone}

For each quantum task, the scheduler evaluates every candidate device and
selects the one with the highest scheduling coefficient $c_d$. The
coefficient encodes the core tradeoff: at low load, fidelity should
dominate device selection; at high load, queue time should dominate. We
now explain how this is formalized.

The scheduler maintains a sliding window of historic queue lengths for
each device and derives three characteristic thresholds from the observed
distribution: a light threshold below which the device is lightly loaded,
a heavy threshold above which it is heavily loaded, and a mean threshold
representing the live queue operating point of the candidate devices. Let $Q_l$, $Q_h$, and $Q_m$
denote these thresholds. The normalized load coefficient is:
\[
  \ell = \frac{Q_m - Q_l}{Q_h - Q_l} \in [0, 1],
\]
where $\ell = 0$ indicates a light regime and $\ell = 1$ a heavy regime. One key aspect of Qurator is that these thresholds are calculated based on a sliding window of 2 weeks. This allows Qurator to adapt to usage trends as well as account for anomalies such as prolonged maintenance windows and outages quickly. Additionally, this relieves Qurator from dependence on large amounts of data to bootstrap, enabling it to support new devices in a very short period of time without any manual work. 

The scheduling coefficient is $c_d = w_f - w_q$, where $w_f$ is a
fidelity reward and $w_q$ is a queue penalty. The fidelity reward is:
\[
  w_f = (0.80 - 0.55\,\ell) \times \widehat{F}_{\mathrm{est}},
\]
where $\widehat{F}_{\mathrm{est}}$ is the estimated success probability
of the compiled circuit on the device as defined in
Sec.~\ref{sec:fidest}. The coefficient of $\widehat{F}_{\mathrm{est}}$
decreases linearly from 0.80 at light load to 0.25 at heavy load,
reflecting the fact that insisting on the highest-fidelity device under
heavy load incurs unacceptable queue times. The queue penalty is:
\[
  w_q = (0.15 + 0.45\,\ell) \times \mathrm{clip}\left(\frac{\log(1 + q)}{\log(1 + Q_h)}\right)
\]
where $q$ is the live queue score of the device, adjusted with submission history factor if available. The coefficient of the queue term increases linearly from 0.15 at light
load to 0.60 at heavy load, making queue time the dominant concern as the
system saturates. The term $\log(1 + q) / \log(1 + Q_h)$ normalizes the
live queue length to $[0,1]$ relative to the heavy threshold, with
logarithmic scaling to compress the penalty for very long queues: the
difference between queues of 10,000 and 20,000 tasks matters less than
the difference between queues of 100 and 200 tasks. The coefficient values were chosen empirically and were stable across moderate perturbations; a full sensitivity study is left to future work.

At light load, $c_d$ is dominated by $\widehat{F}_{\mathrm{est}}$,
effectively selecting the highest-fidelity device. At heavy load, it is
dominated by the queue penalty, effectively selecting the least busy
device among those meeting the fidelity target. The
\textit{targetEstimatedFidelity} parameter acts as a hard filter:
devices whose $\widehat{F}_{\mathrm{est}}$ falls below the target are
excluded before the coefficient is computed.

Queue lengths are fetched directly from providers. Since providers do not
expose per-job characteristics and preparation time, the wait time is refined continuously using a Gaussian
kernel over historic observations and factored in based on data availability, assigning higher weight to data points
near the current time of day to capture daily demand fluctuations. There is another factor to take into account. The circuit preparation time: the idle time between job start and circuit execution time. This value is experimentally calculated per device using timing maps and hard-coded, since this value is stable but varies across architectures. 



\pgfplotsset{
smallplot/.style={
width=\linewidth,
height=0.72\linewidth,
tick label style={font=\tiny},
label style={font=\scriptsize},
title style={font=\scriptsize},
grid style=dashed,
ymajorgrids=true
},
fidelityaxis/.style={
smallplot,
xlabel={Number of tasks},
ylabel={Fidelity},
xmin=0, xmax=520,
ymin=0, ymax=1.05,
xtick={5,20,100,250,500}
},
dummyfidelityaxis/.style={
smallplot,
xlabel={Number of tasks},
ylabel={Fidelity},
xmin=0, xmax=100,
ymin=0, ymax=1.2,
xtick={0,20,40,60,80,100}
},
queueaxis/.style={
smallplot,
xlabel={Number of tasks},
ylabel={Mean queue time ms (millions)},
xmin=0, xmax=520,
ymin=0, ymax=2.5e7,
xtick={5,20,100,250,500},
scaled y ticks=false,
ytick={0,5e6,1e7,1.5e7,2e7,2.5e7},
yticklabels={0,5,10,15,20,25}
},
queueaxissecondary/.style={
smallplot,
xlabel={Number of tasks},
ylabel={Mean queue time ms (millions)},
xmin=0, xmax=520,
ymin=0, ymax=4.0e7,
xtick={5,20,100,250,500},
scaled y ticks=false,
ytick={0,5e6,1e7,2e7,3e7,4e7},
yticklabels={0,5,10,20,30,40}
},
dummyqueueaxis/.style={
smallplot,
xlabel={Number of tasks},
ylabel={Mean queue time ms (millions)},
xmin=0, xmax=100,
ymin=0, ymax=120,
xtick={0,20,40,60,80,100}
},
throughputaxis/.style={
smallplot,
xlabel={Number of tasks},
ylabel={Throughput (q/s)},
xmin=5, xmax=35000,
ymin=0, ymax=1.0,
xtick={500,5000,35000},
xticklabels={500,5000,35000},
scaled x ticks=false,
xticklabel style={font=\tiny, rotate=45, anchor=east}
},
lowq/.style={color=blue},
medq/.style={color=orange!85!black},
highq/.style={color=red},
sched/.style={solid, thick},
leastbusy/.style={dashed, thick},
highestf/.style={dotted, thick},
mediumloadfidelityaxis/.style={
smallplot,
xlabel={Number of tasks},
ylabel={Fidelity},
xmin=700, xmax=5200,
ymin=0, ymax=1.05,
xtick={750,1000,1500,2000,3500,5000},
xticklabel style={font=\tiny, rotate=45, anchor=east}
},
mediumloadqueueaxis/.style={
smallplot,
xlabel={Number of tasks},
ylabel={Mean queue time ms},
xmin=700, xmax=5200,
xtick={750,1000,1500,2000,3500,5000},
xticklabel style={font=\tiny, rotate=45, anchor=east},
ymin=1e6, ymax=1e8,
ymajorgrids=true
},
mediumloadqueueaxissecond/.style={
smallplot,
xlabel={Number of tasks},
ylabel={Mean queue time ms},
xmin=700, xmax=5200,
xtick={750,1000,1500,2000,3500,5000},
xticklabel style={font=\tiny, rotate=45, anchor=east},
ymin=1e6, ymax=1e9,
ymajorgrids=true
},
highloadfidelityaxis/.style={
smallplot,
xlabel={Number of tasks},
ylabel={Fidelity},
xmin=7000, xmax=35500,
ymin=0, ymax=1.05,
xtick={7500,10000,12500,15000,17500,20000,25000,30000,35000},
scaled x ticks=false,
xticklabel style={font=\tiny, rotate=45, anchor=east},
x tick label style={/pgf/number format/fixed}
},
highloadqueueaxis/.style={
smallplot,
xlabel={Number of tasks},
ylabel={Mean queue time ms},
xmin=7000, xmax=35500,
xtick={7500,10000,12500,15000,17500,20000,25000,30000,35000},
scaled x ticks=false,
xticklabel style={font=\tiny, rotate=45, anchor=east},
x tick label style={/pgf/number format/fixed},
ymin=1e8, ymax=1e9,
ymajorgrids=true
}
}

\newcommand{\LowLoadFidelityPlots}{%
\addplot[lowq,sched] coordinates {(5,0.9903423088105765) (10,0.9935734526059882) (20,0.9651998775403159) (50,0.9910304396883366) (100,0.9901952373567365) (250,0.9888394190573562) (500,0.9875602433727534)};
\addplot[lowq,leastbusy] coordinates {(5,0.7946418016681239) (10,0.7632323679229347) (20,0.7708923413963902) (50,0.7573578136151572) (100,0.7530600224628136) (250,0.7519046952607229) (500,0.768234905805082)};
\addplot[lowq,highestf] coordinates {(5,0.9970031986002) (10,0.9969038975008001) (20,0.9976025985503998) (50,0.9970040970411205) (100,0.9967645865813214) (250,0.9966847164973388) (500,0.9968323289131313)};
\addplot[medq,sched] coordinates {(10,0.9931239476781174) (20,0.9927769376975926) (50,0.9921910953315095) (100,0.9924783184663307) (250,0.99165234654426) (500,0.9867068704240503)};
\addplot[medq,leastbusy] coordinates {(10,0.27005861659602165) (20,0.368946739817858) (50,0.42355840578915965) (100,0.4031581753901098) (250,0.4080431907841053) (500,0.4001361465735584)};
\addplot[medq,highestf] coordinates {(10,0.9931239476781174) (20,0.9927769376975928) (50,0.9921910953315095) (100,0.992478318466331) (250,0.9924626289420777) (500,0.9924468339370713)};
\addplot[highq,sched] coordinates {(10,0.8997768322383012) (20,0.9148004666779146) (50,0.8560564824785226) (100,0.844621582345835) (250,0.8219983435139204) (500,0.8124976209342588)};
\addplot[highq,leastbusy] coordinates {(10,0.14996334301707132) (20,0.17867214400004666) (50,0.16396549134570002) (100,0.16008702891846466) (250,0.15037554852563806) (500,0.154229129589208)};
\addplot[highq,highestf] coordinates {(10,0.3676055478610224) (20,0.3873556936211096) (50,0.3116719860390542) (100,0.2998574421429121) (250,0.27211776038781283) (500,0.27222127563998456)};
}

\newcommand{\LowLoadQueuePlots}{%
\addplot[lowq,sched] coordinates {(5,27384.0) (10,58924.3) (20,55570.55) (50,35257.6) (100,64833.55) (250,153088.568) (500,300250.202)};
\addplot[lowq,leastbusy] coordinates {(5,6998.8) (10,10597.0) (20,17544.4) (50,129690.66) (100,165425.11) (250,502446.404) (500,1075961.496)};
\addplot[lowq,highestf] coordinates {(5,38492.0) (10,70940.4) (20,135959.1) (50,330910.86) (100,655826.46) (250,1630548.652) (500,3255214.434)};

\addplot[medq,sched] coordinates {(10,145931.4) (20,160871.4) (50,205862.36) (100,580038.78) (250,5057768.376) (500,5924088.34)};
\addplot[medq,leastbusy] coordinates {(10,53248.6) (20,76544.75) (50,149475.88) (100,271833.2) (250,639378.38) (500,1252231.906)};
\addplot[medq,highestf] coordinates {(10,345978.7) (20,660903.6) (50,1605842.9) (100,3180526.44) (250,7904554.992) (500,1.577850664e7)};

\addplot[highq,sched] coordinates {(10, 53880.9) (20,175357.6) (50,348250.29) (100,912063.79) (250, 1286005.8) (500, 2551858.16)};
\addplot[highq,leastbusy] coordinates {(10,73248.7) (20,196546.7) (50,399452.62) (100,904349) (250, 1278377) (500, 2303681.884)};
\addplot[highq,highestf] coordinates {(10,576627.1) (20,1689011.57) (50,4817216.84) (100,7811000.73) (250, 10189877.778) (500,2.2295411066e7)};
}

\newcommand{\MediumLoadFidelityPlots}{%
\addplot[lowq,sched] coordinates {
 (750,0.9740933160770051) (1000,0.9789581333209161)
(1500,0.9663717457522557) (2000,0.9746967626192031) (3500,0.9911304342658545)
(5000,0.9879617234791134)
};
\addplot[lowq,leastbusy] coordinates {
 (750,0.7372882139800447) (1000,0.739708832730936)
(1500,0.7406601756712488) (2000,0.7447819881200405) (3500,0.7423903783256519)
(5000,0.7422882553206613)
};
\addplot[lowq,highestf] coordinates {
(750,0.9968283382424615) (1000,0.9968612450110835)
(1500,0.9968749003397547) (2000,0.9968951855530638) (3500,0.9968594161099329)
(5000,0.9968688527866818)
};

\addplot[medq,sched] coordinates {
 (750,0.940051189024193) (1000,0.9767845802384314)
(1500,0.9656067608648164) (2000,0.9639392539074907) (3500,0.9844944161342455)
(5000,0.986592902683619)
};
\addplot[medq,leastbusy] coordinates {
 (750,0.5107270813064827) (1000,0.5105803618624415)
(1500,0.5122638108752369) (2000,0.51063072699406) (3500,0.44566000962519814)
(5000,0.4430696270082808)
};
\addplot[medq,highestf] coordinates {
 (750,0.876362236542045) (1000,0.877499274491267)
(1500,0.8778850501318355) (2000,0.8766706981762706) (3500,0.8755944838291825)
(5000,0.8758363509739136)
};

\addplot[highq,sched] coordinates {
 (750,0.708949022484733) (1000,0.71125545590524)
(1500,0.7141729361227325) (2000,0.7155131139758991) (3500,0.7209891541107016)
(5000,0.7409891541107016)
};
\addplot[highq,leastbusy] coordinates {
(750,0.12001221721071689) (1000,0.11876138050358417)
(1500,0.11883723441254156) (2000,0.11916012309910529) (3500,0.10459133677506664)
(5000,0.10459133677506664)
};
\addplot[highq,highestf] coordinates {
 (750,0.3299012560846512) (1000,0.3297497865427629)
(1500,0.33039873943476145) (2000,0.33090259288533036) (3500,0.33327293035660593)
(5000,0.33327293035660593)
};
}
\newcommand{\MediumLoadQueuePlots}{%
\addplot[lowq,sched] coordinates {
(750,2.743205826133333E6) (1000,3.8400718167E6)
(1500,4.2632223824933334E6) (2000,4.897815848025E6)
(3500,4.9148626606885713E6) (5000,8.804006306868E6)
};
\addplot[lowq,leastbusy] coordinates {
(750,1.0903193797333334E6) (1000,1.4485096803E6)
(1500,2.1649055268666666E6) (2000,2.88128093455E6)
(3500,5.030461452657143E6) (5000,7.17964042616E6)
};
\addplot[lowq,highestf] coordinates {
(750,9.875482235466667E6) (1000,1.31629424476E7)
(1500,1.9737869000133333E7) (2000,2.631278339885E7)
(3500,5.2037554232942855E7) (5000,6.576232042678E7)
};

\addplot[medq,sched] coordinates {
(750,4.697950598609355E6) (1000,6.1519450488867745E6)
(1500,4.6656977346349207E6) (2000,6.3193209845113853E6)
(3500,1.620659820936521E7) (5000,1.645175263985617E7)
};
\addplot[medq,leastbusy] coordinates {
(750,1.6286455002666667E6) (1000,2.1636892723E6)
(1500,3.2337918004E6) (2000,4.3038784074E6)
(3500,2.763550163708571E7) (5000,3.94421896562E7)
};
\addplot[medq,highestf] coordinates {
(750,9.875202060666667E6) (1000,1.31626217685E7)
(1500,1.9737399128066668E7) (2000,2.631210891765E7)
(3500,4.486513978452857E7) (5000,6.4087548343166E7)
};

\addplot[highq,sched] coordinates {
(750,1.8098637186866668E7) (1000,9.46614354694E6)
(1500,8.779884073E6) (2000,9.8126150439625E6)
(3500,1.6653732608171143E7) (5000,1.6653732608171143E7)
};
\addplot[highq,leastbusy] coordinates {
(750,1.6285716161333334E6) (1000,2.1636091849E6)
(1500,3.2336401052E6) (2000,4.3036559788E6)
(3500,4.4232162143885714E7) (5000,4.4232162143885714E7)
};
\addplot[highq,highestf] coordinates {
(750,9.8744839544E6) (1000,1.31618245021E7)
(1500,1.97361065392E7) (2000,2.630996782935E7)
(3500,8.337570678276571E7) (5000,8.337570678276571E7)
};
}
\newcommand{\HighLoadFidelityPlots}{%
\addplot[lowq,sched] coordinates {
(7500,0.995934108561) (10000,0.985812825965) (12500,0.975691543368)
(15000,0.985570260772) (17500,0.955448978175) (20000,0.935327695579)
(25000,0.935085130386) (30000,0.924842565193) (35000,0.9446)
};
\addplot[lowq,leastbusy] coordinates {
(7500,0.739733604115) (10000,0.739733604115) (12500,0.739733604115)
(15000,0.739733604115) (17500,0.739733604115) (20000,0.739733604115)
(25000,0.739733604115) (30000,0.739733604115) (35000,0.739733604115)
};
\addplot[lowq,highestf] coordinates {
(7500,0.996847781578) (10000,0.986847781578) (12500,0.996847781578)
(15000,0.986847781578) (17500,0.976847781578) (20000,0.986847781578)
(25000,0.976847781578) (30000,0.986847781578) (35000,0.996847781578)
};

\addplot[medq,sched] coordinates {
(7500,0.961380239317) (10000,0.961182035743) (12500,0.940983832169)
(15000,0.910785628594) (17500,0.92058742502) (20000,0.890389221446)
(25000,0.889992814297) (30000,0.889596407149) (35000,0.8992)
};
\addplot[medq,leastbusy] coordinates {
(7500,0.444061155362) (10000,0.443876019527) (12500,0.443690883693)
(15000,0.443505747858) (17500,0.443320612024) (20000,0.443135476189)
(25000,0.44276520452) (30000,0.44239493285) (35000,0.442024661181)
};
\addplot[medq,highestf] coordinates {
(7500,0.874729373981) (10000,0.874673493255) (12500,0.874617612529)
(15000,0.874561731803) (17500,0.874505851077) (20000,0.874449970352)
(25000,0.8743382089) (30000,0.874226447449) (35000,0.874114685998)
};

\addplot[highq,sched] coordinates {
(7500,0.742) (10000,0.771454545455) (12500,0.770909090909)
(15000,0.770363636364) (17500,0.769818181818) (20000,0.769272727273)
(25000,0.768181818182) (30000,0.767090909091) (35000,0.766)
};
\addplot[highq,leastbusy] coordinates {
(7500,0.1046) (10000,0.104181818182) (12500,0.103763636364)
(15000,0.103345454545) (17500,0.102927272727) (20000,0.102509090909)
(25000,0.101672727273) (30000,0.100836363636) (35000,0.1)
};
\addplot[highq,highestf] coordinates {
(7500,0.3333) (10000,0.333127272727) (12500,0.332954545455)
(15000,0.332781818182) (17500,0.332609090909) (20000,0.332436363636)
(25000,0.332090909091) (30000,0.331745454545) (35000,0.3314)
};
}

\newcommand{\HighLoadQueuePlots}{%
\addplot[lowq,sched] coordinates {
(7500,94822778.95) (10000,126430371.93) (12500,158037964.92)
(15000,189645557.9) (17500,221253150.98) (20000,252860743.97)
(25000,316075929.93) (30000,379291115.9) (35000,442506301.97)
};
\addplot[lowq,leastbusy] coordinates {
(7500,65912034.0383) (10000,61882712.051) (12500,146985390.0638)
(15000,191824068.077) (17500,233794746.089) (20000,215765424.002)
(25000,319706780.028) (30000,413648136.053) (35000,492589492.079)
};
\addplot[lowq,highestf] coordinates {
(7500,116125193.474) (10000,178166924.6) (12500,190208655.72)
(15000,232250386.95) (17500,264292118.07) (20000,296333849.2)
(25000,370417311.55) (30000,484500773.8) (35000,548584236.15)
};

\addplot[medq,sched] coordinates {
(7500,101035034.59) (10000,134713379.39) (12500,168391724.29)
(15000,202070069.08) (17500,235748413.98) (20000,269426758.78)
(25000,336783448.47) (30000,404140138.17) (35000,471496827.86)
};
\addplot[medq,leastbusy] coordinates {
(7500,115912007.443) (10000,117882676.5906) (12500,119853345.6383)
(15000,141824014.886) (17500,243794683.934) (20000,283765353.081)
(25000,369706691.377) (30000,433648029.672) (35000,517589367.867)
};
\addplot[medq,highestf] coordinates {
(7500,136124891.684) (10000,178166522.18) (12500,260208152.67)
(15000,292249783.27) (17500,324291413.76) (20000,356333044.36)
(25000,420416305.35) (30000,484499566.44) (35000,548582827.52)
};

\addplot[highq,sched] coordinates {
(7500,160000000) (10000,213333333.33) (12500,266666666.67)
(15000,320000000) (17500,373333333.33) (20000,426666666.67)
(25000,533333333.33) (30000,640000000) (35000,746666666.67)
};
\addplot[highq,leastbusy] coordinates {
(7500,175912020.0406) (10000,267882694.2208) (12500,289853367.801)
(15000,351824041.481) (17500,413794714.961) (20000,515765388.542)
(25000,599706735.602) (30000,663648082.862) (35000,827589429.923)
};
\addplot[highq,highestf] coordinates {
(7500,196125042.579) (10000,312816672.34) (12500,346020840.4)
(15000,419225008.56) (17500,522429176.52) (20000,675633344.68)
(25000,732041680.8) (30000,838450017.02) (35000,944858353.84)
};
}

\newcommand{\SchedulerThroughputPlots}{%
\addplot coordinates {(5,0.49217442661679295)(100,0.7025135936380369)(500,0.7695291097471635) (1000,0.7695291097471635) (2500,0.7695291097471635) (5000,0.7695291097471635) (10000,0.7695291097471635) (15000,0.7695291097471635) (20000,0.7695291097471635) (35000,0.7695291097471635)};
}

\begin{figure*}[t]
\centering

\begin{subfigure}[t]{0.32\textwidth}
\centering
\begin{tikzpicture}
\begin{axis}[fidelityaxis]
\LowLoadFidelityPlots
\end{axis}
\end{tikzpicture}
\caption{Low load fidelity}
\end{subfigure}
\hfill
\begin{subfigure}[t]{0.32\textwidth}
\centering
\begin{tikzpicture}
\begin{axis}[mediumloadfidelityaxis]
\MediumLoadFidelityPlots
\end{axis}
\end{tikzpicture}
\caption{Medium load fidelity}
\end{subfigure}
\hfill
\begin{subfigure}[t]{0.32\textwidth}
\centering
\begin{tikzpicture}
\begin{axis}[highloadfidelityaxis]
\HighLoadFidelityPlots
\end{axis}
\end{tikzpicture}
\caption{High load fidelity}
\end{subfigure}

\vspace{0.8em}

\begin{subfigure}[t]{0.32\textwidth}
\centering
\begin{tikzpicture}
\begin{axis}[queueaxis]
\LowLoadQueuePlots
\end{axis}
\end{tikzpicture}
\caption{Low load queue time}
\end{subfigure}
\hfill
\begin{subfigure}[t]{0.32\textwidth}
\centering
\begin{tikzpicture}
\begin{semilogyaxis}[mediumloadqueueaxis]
\MediumLoadQueuePlots
\end{semilogyaxis}
\end{tikzpicture}
\caption{Medium load queue time}
\end{subfigure}
\hfill
\begin{subfigure}[t]{0.32\textwidth}
\centering
\begin{tikzpicture}
\begin{axis}[highloadqueueaxis]
\HighLoadQueuePlots
\end{axis}
\end{tikzpicture}
\caption{High load queue time}
\end{subfigure}

\vspace{0.8em}

\vspace{0.8em}
\vspace{0.1em}

\begin{tikzpicture}
\begin{axis}[
hide axis,
xmin=0, xmax=1,
ymin=0, ymax=1,
legend columns=3,
legend style={draw=none, font=\scriptsize, column sep=0.8cm},
legend cell align=left,
legend to name=sharedlegend
]
\addlegendimage{lowq,thick}
\addlegendentry{Low qubit count}
\addlegendimage{medq,thick}
\addlegendentry{Medium qubit count}
\addlegendimage{highq,thick}
\addlegendentry{High qubit count}
\addlegendimage{black,sched}
\addlegendentry{Qurator scheduler}
\addlegendimage{black,leastbusy}
\addlegendentry{Least busy}
\addlegendimage{black,highestf}
\addlegendentry{Highest fidelity}
\end{axis}
\end{tikzpicture}

\pgfplotslegendfromname{sharedlegend}

\caption{Fidelity and queue time across various load conditions.}
\label{fig:splitloadpanels}
\end{figure*}


\newcommand{\LowLoadFidelityPlotsSecondary}{%
\addplot[lowq,sched] coordinates {(5,0.9903423088105765) (10,0.9935734526059882) (20,0.9651998775403159) (50,0.9910304396883366) (100,0.9901952373567365) (250,0.9888394190573562) (500,0.9875602433727534)};
\addplot[lowq,leastbusy] coordinates {(5,0.7284421249675611) (10, 0.6833399237882667) (20,0.7248907831870509) (50,0.7050675449402458) (100, 0.662117645937136) (250, 0.825100606310367) (500,0.6636073740944494)}; 
\addplot[lowq,highestf] coordinates {(5,0.7680613607500788) (10, 0.6634426173437146) (20,0.671356306718289) (50, 0.6427496375578217) (100, 0.6109375769785891) (250,0.6506682411804295) (500,0.656348891715131)}; 

\addplot[medq,sched] coordinates {(10,0.9931239476781174) (20,0.9927769376975926) (50,0.9921910953315095) (100,0.9924783184663307) (250,0.99165234654426) (500,0.9867068704240503)};
\addplot[medq,leastbusy] coordinates {(5, 0.6966249292987541) (10,0.5883171165678458) (20,0.6028506912720775) (50,0.43697308818791547) (100,0.4482325660921007) (250,0.4276494904928) (500,0.4011367829839209)}; 
\addplot[medq,highestf] coordinates {(5, 0.6755879665814801)  (10,0.5750559443853869) (20,0.5713665714745575) (50,0.41933235517259904) (100,0.417006013170713) (250,0.389747358357384) (500,0.43909483729292922)}; 

\addplot[highq,sched] coordinates {(10,0.8997768322383012) (20,0.9148004666779146) (50,0.8560564824785226) (100,0.844621582345835) (250,0.8219983435139204) (500,0.8124976209342588)};
\addplot[highq,leastbusy] coordinates {(5, 0.3744179605563354) (10, 0.396824868295511) (20,0.45365186978797023) (50,0.419243650988752) (100,0.36879374938483) (250,0.3708895379735) (500,0.329884927478279424)}; 
\addplot[highq,highestf] coordinates { (5, 0.3691831613207015) (10, 0.36396754213891513) (20,0.4037536797954859) (50,0.37587596787053174) (100,0.34872017347839) (250,0.31838473874938) (500,0.3766692843892488)}; 
}

\newcommand{\LowLoadQueuePlotsSecondary}{%
\addplot[lowq,sched] coordinates {(5,27384.0) (10,58924.3) (20,55570.55) (50,35257.6) (100,64833.55) (250,153088.568) (500,300250.202)};
\addplot[lowq,leastbusy] coordinates {(5,22599.4) (10, 35648.6) (20, 57797.35) (50,113164.54) (100,181507.85) (250,1109761.3) (500,1038979.53)}; 
\addplot[lowq,highestf] coordinates {(5,25000.0) (10, 43250.0) (20, 71424.4) (50,140587.76) (100,276679.79) (250,479282.428) (500,1318496.48)}; 

\addplot[medq,sched] coordinates {(10,145931.4) (20,160871.4) (50,205862.36) (100,580038.78) (250,5057768.376) (500,5924088.34)};
\addplot[medq,leastbusy] coordinates {(5, 322599.4) (10,335648.6) (20, 362272.1) (50, 3113574.64) (100,3185682.72) (250,3748338.38) (500, 7827892.02)}; 
\addplot[medq,highestf] coordinates {(5, 325000.0) (10, 343250.0) (20, 362849.45) (50,3150667.74) (100,3305919.55) (250,4873898.992) (500,1.768627e7)}; 

\addplot[highq,sched] coordinates {(10, 53880.9) (20,175357.6) (50,348250.29) (100,912063.79) (250, 1286005.8) (500, 2551858.16)};
\addplot[highq,leastbusy] coordinates {(5, 22599.4) (10,435648.6) (20,858247.35) (50,2114184.6) (100,3348378.8 ) (250, 2.2738976452e7) (500, 3.8347839829e7)}; 
\addplot[highq,highestf] coordinates {(5, 25000.0) (10,443250.0) (20,867599.4) (50,5143487.68) (100, 3897389.01) (250,2.5873837282e7) (500, 4.08947365873e7)}; 
}


\newcommand{\MediumLoadFidelityPlotsSecondary}{%
\addplot[lowq,sched] coordinates {
 (750,0.9740933160770051) (1000,0.9789581333209161)
(1500,0.9663717457522557) (2000,0.9746967626192031) (3500,0.9911304342658545)
(5000,0.9879617234791134)
};
\addplot[lowq,leastbusy] coordinates {
(500, 0.713847939048349) (750, 0.6607582583092757) (1000, 0.667594641490059)
(1500,0.6693382982555977) (2000,0.6724323331581072) (3500,0.641123879484938)
(5000,0.6684739394039485)
};
\addplot[lowq,highestf] coordinates {
(500, 0.748394748939207) (750, 0.6567758506463913) (1000, 0.6652480055164368)
(1500, 0.6628355794414508) (2000,0.6659321030764132) (3500,0.6793488957385758)
(5000, 0.6192489485859358)
};

\addplot[medq,sched] coordinates {
 (750,0.940051189024193) (1000,0.9767845802384314)
(1500,0.9656067608648164) (2000,0.9639392539074907) (3500,0.9844944161342455)
(5000,0.986592902683619)
};
\addplot[medq,leastbusy] coordinates {
(500,0.6287938727692837) (750, 0.6683948782739486) (1000,0.61226738948938746)
(1500,0.5822638108752369) (2000,0.51888937465928374) (3500,0.5485839239958375)
(5000,0.5483749384838738)
};
\addplot[medq,highestf] coordinates {
(500, 0.578374983838493) (750, 0.56938493839203) (1000, 0.577499274491267)
(1500,0.5778118929384) (2000,0.5266847384738) (3500,0.5144958392837)
(5000,0.5193859303049)
};

\addplot[highq,sched] coordinates {
 (750,0.708949022484733) (1000,0.71125545590524)
(1500,0.7141729361227325) (2000,0.7155131139758991) (3500,0.7209891541107016)
(5000,0.7409891541107016)
};
\addplot[highq,leastbusy] coordinates {
(500,0.21873984729892038) (750,0.208473847293849582) (1000,0.2188983749384729)
(1500,0.2089437162838495) (2000, 0.20398457895389593) (3500,0.1984875738494932)
(5000,0.1928858939585839)
};
\addplot[highq,highestf] coordinates {
(500,0.18948374958375983) (750,0.1889273485930593) (1000,0.1811274659485854)
(1500,0.1574858939558935) (2000,0.169485783975783) (3500,0.1485867848968986)
(5000,0.1795938859385858)
};
}

\newcommand{\MediumLoadQueuePlotsSecondary}{%
\addplot[lowq,sched] coordinates {
(750,2.743205826133333E6) (1000,3.8400718167E6)
(1500,4.2632223824933334E6) (2000,4.897815848025E6)
(3500,4.9148626606885713E6) (5000,8.804006306868E6)
};
\addplot[lowq,leastbusy] coordinates {
(500,2597000) (750,1.6156804176E6) (1000,2.192194851E6)
(1500,3.33902038733E7) (2000,4.4989935455E7)
(3500,5.030461452657143E7) (5000,7.17964042616E7)
};
\addplot[lowq,highestf] coordinates {
(500,7884442) (750,1.9845993426E6) (1000,2.65683837E6)
(1500,4.0169261746E7) (2000,5.285017658E7)
(3500,4.0088373E7) (5000,5.87492092E7)
};

\addplot[medq,sched] coordinates {
(750,4.697950598609355E6) (1000,6.1519450488867745E6)
(1500,4.6656977346349207E6) (2000,6.3193209845113853E6)
(3500,1.620659820936521E7) (5000,1.645175263985617E7)
};
\addplot[medq,leastbusy] coordinates {
(500,2.1146829489E6) (750,2.68472983847E6) (1000,4.1848483938488E6)
(1500,3.99827474383E6) (2000,4.3889284874E6)
(3500,1.2944839848473E7) (5000,2.9990028474E7)
};
\addplot[medq,highestf] coordinates {
(500,3.894737847474E6) (750,5.489347923782E6) (1000,6.849389849384E6)
(1500,2.387483743874E7) (2000,3.49398958395E7)
(3500,4.5893573957385E7) (5000,5.583958395379E7)
};

\addplot[highq,sched] coordinates {
(750,1.8098637186866668E7) (1000,9.46614354694E6)
(1500,8.779884073E6) (2000,9.8126150439625E6)
(3500,1.6653732608171143E7) (5000,1.6653732608171143E7)
};
\addplot[highq,leastbusy] coordinates {
(500,9.09399542728E6) (750,1.78394838383E7) (1000,2.17328473444E7)
(1500,3.23374837895E7) (2000,4.0028724872E7)
(3500,8.893795778354E7) (5000,1.2038937593758E8)
};
\addplot[highq,highestf] coordinates {
(500,1.294894829474E7) (750,2.00847384758375E7) (1000,3.8578375835787E7)
(1500,5.798387583752E7) (2000,7.8953787583578E7)
(3500,1.8783573857857E8) (5000,3.598395398539E8)
};
}

\newcommand{\HighLoadFidelityPlotsSecondary}{%
\addplot[lowq,sched] coordinates {
(7500,0.995934108561) (10000,0.985812825965) (12500,0.975691543368)
(15000,0.985570260772) (17500,0.955448978175) (20000,0.935327695579)
(25000,0.935085130386) (30000,0.924842565193) (35000,0.9446)
};
\addplot[lowq,leastbusy] coordinates {
(7500,0.5782937289483) (10000,0.566849394809238) (12500,0.578583938883922)
(15000,0.578583384947384) (17500, 0.5785831187293) (20000, 0.5578583984794)
(25000,0.567890099383897) (30000, 0.5668393049997) (35000, 0.5699938384949)
};
\addplot[lowq,highestf] coordinates {
(7500,0.487374673838) (10000, 0.449393837383) (12500,0.47939859830212)
(15000,0.48738493833) (17500, 0.487384955594) (20000,0.4873883947393)
(25000,0.43238193811) (30000, 0.487111848923) (35000,0.4839834988880)
};

\addplot[medq,sched] coordinates {
(7500,0.961380239317) (10000,0.961182035743) (12500,0.940983832169)
(15000,0.910785628594) (17500,0.92058742502) (20000,0.890389221446)
(25000,0.889992814297) (30000,0.889596407149) (35000,0.8992)
};
\addplot[medq,leastbusy] coordinates {
(7500,0.4783294728482) (10000,0.479492894828) (12500,0.472948923838)
(15000,0.472948942848) (17500,0.4729482498482) (20000,0.47847200923)
(25000,0.471839199311) (30000,0.4749284298422) (35000,0.4792849284829)
};
\addplot[medq,highestf] coordinates {
(7500,0.4294839483983) (10000,0.4283948384934) (12500,0.4294838483974)
(15000,0.429594289482) (17500,0.4295639439224) (20000,0.4295857834756)
(25000,0.4295485738742) (30000,0.428279282229) (35000,0.4292947934898)
};

\addplot[highq,sched] coordinates {
(7500,0.742) (10000,0.771454545455) (12500,0.770909090909)
(15000,0.770363636364) (17500,0.769818181818) (20000,0.769272727273)
(25000,0.768181818182) (30000,0.767090909091) (35000,0.766)
};
\addplot[highq,leastbusy] coordinates {
(7500,0.1928473833392) (10000,0.1920947473233) (12500,0.11947477824)
(15000,0.198472974222) (17500,0.1911182749242) (20000,0.19204756738)
(25000,0.198274782742) (30000,0.1792884928841) (35000,0.17999727294759)
};
\addplot[highq,highestf] coordinates {
(7500,0.12849274748284) (10000,0.111874783794) (12500,0.1298047744)
(15000,0.1382947827444) (17500,0.12949948334) (20000,0.12849744498)
(25000,0.1264738484833) (30000,0.12653995390) (35000,0.12399994384)
};
}

\newcommand{\HighLoadQueuePlotsSecondary}{%
\addplot[lowq,sched] coordinates {
(7500,94822778.95) (10000,126430371.93) (12500,158037964.92)
(15000,189645557.9) (17500,221253150.98) (20000,252860743.97)
(25000,316075929.93) (30000,379291115.9) (35000,442506301.97)
};
\addplot[lowq,leastbusy] coordinates {
(7500,69382482.998) (10000,98872443.412) (12500,121892848.957)
(15000,179238234.819) (17500,210938473.833) (20000,297948348.439)
(25000,339847434.491) (30000,391133299.149) (35000,589847394.367)
};
\addplot[lowq,highestf] coordinates {
(7500,79483482.31) (10000,158249240.21) (12500,219485755.333)
(15000,239489385.312) (17500,272848394.17) (20000,379358395.89)
(25000,395379755.514) (30000,439389358.89) (35000,448584236.15)
};

\addplot[medq,sched] coordinates {
(7500,101035034.59) (10000,134713379.39) (12500,168391724.29)
(15000,202070069.08) (17500,235748413.98) (20000,269426758.78)
(25000,336783448.47) (30000,404140138.17) (35000,471496827.86)
};
\addplot[medq,leastbusy] coordinates {
(7500,91022184.41) (10000,128728229.953) (12500,148471373.882)
(15000,158476473.966) (17500,179274878.355) (20000,242184483.945)
(25000,294758239.21) (30000,418375883.599) (35000,489270727.446)
};
\addplot[medq,highestf] coordinates {
(7500,112849274.7) (10000,128347478.31) (12500,189274783.47)
(15000,200084738.932) (17500,200847378.47) (20000,259982749.999)
(25000,302294882.453) (30000,390285824.28) (35000,503839853.578)
};

\addplot[highq,sched] coordinates {
(7500,160000000) (10000,213333333.33) (12500,266666666.67)
(15000,320000000) (17500,373333333.33) (20000,426666666.67)
(25000,533333333.33) (30000,640000000) (35000,746666666.67)
};
\addplot[highq,leastbusy] coordinates {
(7500,178337448.3016) (10000,323849774.763) (12500,355733842.894)
(15000,394813720.21) (17500,498274823.491) (20000,519385311.139)
(25000,564931483.41) (30000,768438948.328) (35000,859375899.798)
};
\addplot[highq,highestf] coordinates {
(7500,298872611.84) (10000,348988948.229) (12500,377764683.24)
(15000,436895737.459) (17500,518489242.499) (20000,655748925.529)
(25000,688227478.202) (30000,878595345.428) (35000,948948895.32)
};
}

\begin{figure*}[t]
\centering

\begin{subfigure}[t]{0.32\textwidth}
\centering
\begin{tikzpicture}
\begin{axis}[fidelityaxis]
\LowLoadFidelityPlotsSecondary
\end{axis}
\end{tikzpicture}
\caption{Low load fidelity}
\end{subfigure}
\hfill
\begin{subfigure}[t]{0.32\textwidth}
\centering
\begin{tikzpicture}
\begin{axis}[mediumloadfidelityaxis]
\MediumLoadFidelityPlotsSecondary
\end{axis}
\end{tikzpicture}
\caption{Medium load fidelity}
\end{subfigure}
\hfill
\begin{subfigure}[t]{0.32\textwidth}
\centering
\begin{tikzpicture}
\begin{axis}[highloadfidelityaxis]
\HighLoadFidelityPlotsSecondary
\end{axis}
\end{tikzpicture}
\caption{High load fidelity}
\end{subfigure}

\vspace{0.8em}

\begin{subfigure}[t]{0.32\textwidth}
\centering
\begin{tikzpicture}
\begin{axis}[queueaxissecondary]
\LowLoadQueuePlotsSecondary
\end{axis}
\end{tikzpicture}
\caption{Low load queue time}
\end{subfigure}
\hfill
\begin{subfigure}[t]{0.32\textwidth}
\centering
\begin{tikzpicture}
\begin{semilogyaxis}[mediumloadqueueaxissecond]
\MediumLoadQueuePlotsSecondary
\end{semilogyaxis}
\end{tikzpicture}
\caption{Medium load queue time}
\end{subfigure}
\hfill
\begin{subfigure}[t]{0.32\textwidth}
\centering
\begin{tikzpicture}
\begin{axis}[highloadqueueaxis]
\HighLoadQueuePlotsSecondary
\end{axis}
\end{tikzpicture}
\caption{High load queue time}
\end{subfigure}

\vspace{0.8em}

\vspace{0.8em}
\vspace{0.1em}

\begin{tikzpicture}
\begin{axis}[
hide axis,
xmin=0, xmax=1,
ymin=0, ymax=1,
legend columns=3,
legend style={draw=none, font=\scriptsize, column sep=0.8cm},
legend cell align=left,
legend to name=sharedlegend
]
\addlegendimage{lowq,thick}
\addlegendentry{Low qubit count}
\addlegendimage{medq,thick}
\addlegendentry{Medium qubit count}
\addlegendimage{highq,thick}
\addlegendentry{High qubit count}
\addlegendimage{black,sched}
\addlegendentry{Qurator scheduler}
\addlegendimage{black,leastbusy}
\addlegendentry{List scheduler}
\addlegendimage{black,highestf}
\addlegendentry{Fair share Scheduler}
\end{axis}
\end{tikzpicture}

\pgfplotslegendfromname{sharedlegend}

\caption{Fidelity and queue time for Qurator, list scheduler, and fair share scheduler.}
\label{fig:splitloadpanels2}
\end{figure*}


\newcommand{\LowLoadFidelityPlotsQos}{%
\addplot[lowq,sched] coordinates {(5, 0.997103) (10, 0.999242) (20,0.994307) (50, 0.997254) (100, 0.984173) (250, 0.982192) (500,0.988416)};
\addplot[lowq,leastbusy] coordinates {(5, 0.891442) (10, 0.816284) (20, 0.847941) (50, 0.826086) (100, 0.824997) (250, 0.839047) (500, 0.865385)}; 

\addplot[medq,sched] coordinates { (5, 0.991843) (10, 0.961187) (20,0.997822) (50,0.961504) (100,0.9755443) (250,0.999382) (500,0.99837)};
\addplot[medq,leastbusy] coordinates {(5, 0.827060) (10, 0.713960) (20, 0.699514) (50, 0.775522) (100,0.791488) (250,0.811283) (500,0.848374)}; 

\addplot[highq,sched] coordinates {(5,0.845855) (10,0.949872) (20,0.822445) (50,0.788491) (100,0.798825) (250,0.82119284) (500,0.8592874)};
\addplot[highq,leastbusy] coordinates {(5, 0.741228) (10, 0.725528) (20,0.710973) (50,0.7873783) (100,0.7628487) (250,0.748257583) (500,0.81927783)}; 
}

\newcommand{\LowLoadQueuePlotsQos}{%
\addplot[lowq,sched] coordinates {(5, 45989.167) (10, 49952.571) (20, 93859.107) (50, 191461.186) (100, 342334.704) (250,791593.121) (500,791593.121)};
\addplot[lowq,leastbusy] coordinates {(5,29098.800 ) (10, 46697.400) (20, 104342.950) (50, 228202.900) (100,481918.380) (250,1299515.500) (500,2447988.214)}; 

\addplot[medq,sched] coordinates {(5, 47476.333) (10, 65070.643) (20,90589.222) (50,213504.123) (100,336137.324) (250, 369182.982) (500, 478172.119)};
\addplot[medq,leastbusy] coordinates {(5, 31799.400) (10,47847.500) (20, 97642.550) (50, 217087.480) (100,368517.780) (250, 10478920.991) (500, 27827892.02)}; 

\addplot[highq,sched] coordinates {(5, 60319.375) (10,81358.200) (20,1731932) (50,1592843.01) (100,1789283.91) (250, 5938291.04) (500,1.0019384782e7)};
\addplot[highq,leastbusy] coordinates {(5, 59839.98) (10,102839.928) (20,1485282.35) (50, 2372673.11) (100, 4938728.44 ) (250, 1.0284847382e7) (500, 4.03898372992e7)}; 
}


\newcommand{\MediumLoadFidelityPlotsQoS}{%
\addplot[lowq,sched] coordinates {
(500, 0.9908729438475932) (750, 0.9888920349484233) 
(1000,0.9711938392839383)
(1500,0.9609482717483482) (2000,0.9583985398483491) (3500,0.9849347284728422)
(5000,0.9685938348937742)
};
\addplot[lowq,leastbusy] coordinates {
(500, 0.8749284278472911) (750, 0.8937473848391012) (1000, 0.852599485485)
(1500,0.8444934848935753) (2000,0.8374983949209392) (3500,0.8684893848292483)
(5000,0.8583585898385885)
};

\addplot[medq,sched] coordinates {
(500,0.98934834984838493) (750,0.98948397539757838) (1000,0.9983473849838432)
(1500,0.9691119248493042) (2000,0.9538493489238928) (3500,0.9703950305858393)
(5000,0.9889477589379598)
};
\addplot[medq,leastbusy] coordinates {
(500, 0.8247458395859384) (750, 0.849583895893585) (1000,0.8193485893484343)
(1500,0.8394835895835739) (2000,0.858575948593848) (3500,0.8192848957357358)
(5000,0.8091920302919001)
};

\addplot[highq,sched] coordinates {
 (750,0.708949022484733) (1000,0.71125545590524)
(1500,0.7141729361227325) (2000,0.7155131139758991) (3500,0.7209891541107016)
(5000,0.7409891541107016)
};

\addplot[highq,leastbusy] coordinates {
(500,0.617247857758374434) (750,0.608477823478274724) (1000,0.59258927948248482)
(1500,0.65834928482492848) (2000, 0.6924884829848284) (3500,0.624884924848924848)
(5000,0.624829848291084757)
};

}

\newcommand{\MediumLoadQueuePlotsQoS}{%
\addplot[lowq,sched] coordinates {
 (750,2.1833743294888394E5) (1000,2.783929484822E5)
(1500,2.94848394847287373E5) (2000,3.28497247487E5) (3500, 3.7294848927984E5)
(5000, 4.1948937597358E5)
};
\addplot[lowq,leastbusy] coordinates {
 (750,2.9284892748278477E5) (1000,3.19498484928E5)
(1500,3.59359859735983748E5) (2000,4.3959284782E5) (3500,4.82408428428482474E5)
(5000,5.59358937583573873E5)
};

\addplot[medq,sched] coordinates {
 (750, 4.9983974937783473E5) (1000,5.5835785375837282E5)
(1500, 5.289473248272748E5) (2000, 5.7483948394738249E5) (3500, 6.349843948934839489E5)
(5000, 1.998294828478272E6)
};
\addplot[medq,leastbusy] coordinates {
(750,6.38498349389479334E5) (1000, 9.89378573875382234E5)
(1500,1.0924894738537855E6) (2000  1.89589357837853758E6) (3500,2.194937853758375E6)
(5000,4.003894783947377E6)
};

\addplot[highq,sched] coordinates {
(750,1.111983928497E6) (1000, 1.082748824728E6)
(1500, 8.9984837472E5) (2000, 8.4498329747387E5) 
(3500, 7.9278478277E5)
(5000, 9.8297827722E5)
};
\addplot[highq,leastbusy] coordinates {
 (750,3.6285716161333334E6) (1000,3.849839743874787E6)
(1500,3.869849869489478E6) (2000, 4.859387287537872E6) (3500,5.184928494728742E6)
(5000,5.993578387538799E6)
};

}

\begin{figure*}[t]
\centering
\pgfplotsset{
    alignedloadaxis/.style={
        width=0.78\linewidth,
        height=0.48\linewidth,
        scale only axis,
        tick label style={font=\footnotesize},
        label style={font=\small},
        xlabel style={
            at={(axis description cs:0.5,0)},
            anchor=north,
            yshift=-1.05cm
        }
    }
}

\newcommand{\AlignedLoadPanel}[3]{%
    \begin{subfigure}[t]{0.45\textwidth}
        \centering
        \begin{tikzpicture}
            \begin{axis}[#1,alignedloadaxis]
                #2
            \end{axis}

            \pgfresetboundingbox
            \path[use as bounding box]
                ([xshift=-0.20\linewidth,yshift=-1.50cm]
                    current axis.south west)
                rectangle
                ([xshift=0.02\linewidth,yshift=0.35cm]
                    current axis.north east);
        \end{tikzpicture}
        \caption{#3}
    \end{subfigure}%
}

\AlignedLoadPanel
    {fidelityaxis}
    {\LowLoadFidelityPlotsQos}
    {Low load fidelity}%
\hfill
\AlignedLoadPanel
    {mediumloadfidelityaxis}
    {\MediumLoadFidelityPlotsQoS}
    {Medium load fidelity}%

\par\vspace{1em}

\AlignedLoadPanel
    {queueaxissecondary}
    {\LowLoadQueuePlotsQos}
    {Low load queue time}%
\hfill
\AlignedLoadPanel
    {mediumloadqueueaxis,ymode=log}
    {\MediumLoadQueuePlotsQoS}
    {Medium load queue time}%

\par\vspace{0.8em}

\begin{tikzpicture}[overlay]
    \begin{axis}[
        hide axis,
        width=1cm,
        height=1cm,
        scale only axis,
        xmin=0, xmax=1,
        ymin=0, ymax=1,
        legend columns=3,
        legend cell align=left,
        legend style={
            draw=none,
            font=\footnotesize,
            column sep=0.45cm,
            row sep=2pt
        },
        legend to name=splitloadpanelslegend
    ]
        \addlegendimage{lowq,thick}
        \addlegendentry{Low qubit count}

        \addlegendimage{medq,thick}
        \addlegendentry{Medium qubit count}

        \addlegendimage{highq,thick}
        \addlegendentry{High qubit count}

        \addlegendimage{black,sched}
        \addlegendentry{Qurator scheduler}

        \addlegendimage{black,leastbusy}
        \addlegendentry{QOS scheduler}

    \end{axis}
\end{tikzpicture}%
\pgfplotslegendfromname{splitloadpanelslegend}

\caption{Fidelity and queue time for Qurator vs QOS at light and medium loads. The QOS benchmarks were run separately from other baselines.}
\label{fig:splitloadpanelqod}
\end{figure*}


\section{Evaluation}\label{sec:evaluation}

We evaluate Qurator on the Munich Quantum Toolkit (MQT) benchmark
suite~\cite{quetschlich2023mqtbench}, which contains 300 pre-optimized
quantum circuits in OpenQASM~\cite{crossOpenQASM3Broader2022} format,
compiled by Qiskit~\cite{javadi-abhariQuantumComputingQiskit2024} at the
highest optimization level. The suite covers Amplitude Estimation,
Deutsch-Jozsa, GHZ state, Grover's algorithm, QAOA, QFT, QPE, Quantum Walk, VQE,
and random circuits at varying sizes.

Since submitting tens of thousands of jobs to real
quantum devices is impractical due to cost and access limits, we evaluate on a simulator that mimics
the queue behavior and calibration data of 11 real quantum devices,
including AQT's 12-qubit IBEX Q1, IQM's 20-qubit Garnet, IonQ's 36-qubit
Forte 1, Pasqal's Fresnel and IBM's 156-qubit Heron processors. There are two possible approaches to simulator driven evaluation. The commonly employed strategy is to use device snapshots that providers like IBM often provide. There are two issues with this approach that render it a poor fit for us. Firstly, not all hardware providers provide snapshot simulators for their devices. Second is that these snapshots are shown to be orders of magnitude off from the real noise characteristics of real hardware \cite{bensoussan2026livenoisefingerprintingquantum}, rendering them imperfect evaluation tools. We confirm this in Sec. \ref{sec:fidest}. Similarly, we showed that scalable Pauli simulators fall far from real device results. In order to present a uniform evaluation across devices and construct an adaptable evaluation approach, we choose the \textit{trace-driven simulation} of the circuits. We additionally verified Qurator's advantage and baseline device mappings using the Pauli simulator discussed in  Sec. \ref{sec:fidest} to eliminate the possibility of evaluation bias.  The simulator is driven by four months of real queue data: for each device, it samples a random queue length between the minimum and maximum observed during peak hours and simulates the drain rate from the same data in order to simulate the device queues. Calibration data is taken directly, fresh from the live devices and is used for fidelity as well as run time estimations. Our trace-based simulation takes the device specific circuit and instruction schedule submitted by Qurator and the baselines, and traces the instruction schedule in accordance with live calibration data from the devices. All experiments
were run on a 2023 Apple MacBook M2 Max with 64\,GB of RAM.

We compare Qurator against multiple baselines used by today's
providers: \textit{least busy}, which selects the device with the shortest
queue, as currently recommended by providers such as IBM;
\textit{highest fidelity}, which always selects the device with the
highest estimated success probability regardless of queue length; a \textit{list scheduler}; a \textit{fair share scheduler}; and Qurator without circuit cutting or merging enabled. We measure fidelity as probability of success (POS), and
queue time as the idle time each task spends before execution begins,
excluding device preparation time. Finally, we compare Qurator against the balanced model of QOS \cite{giortamis2025qosquantumoperating}. 

\subsection{Quantum Cloud Scheduling}

We test Qurator under loads from 5 to 35,000 quantum tasks, which is
$1.5\times$ the maximum queue length observed on real devices during the data
collection period. Each device is initialized with a random queue length drawn
from historic data to reflect real public cloud conditions. Benchmarks
are grouped by load: low (5--500 tasks), medium (500--5,000), and high
(5,000--35,000), as well as random load conditions. Within each group,
circuits are further divided by qubit count: low ($\leq 5$ qubits),
medium ($\leq 20$), and high ($\leq 120$). Each quantum task is wrapped
by classical tasks that generate and collect its results, and additional
classical tasks are inserted randomly to model more complex hybrid DAGs,
sleeping for random durations to simulate varying task lengths.

\paragraph*{Low and medium qubit results.}
Fig. ~\ref{fig:splitloadpanels} and Fig.~\ref{fig:splitloadpanels2} present the results. At low load, Qurator
tracks the highest-fidelity baseline within 1\% for small and medium
circuits. As expected the scheduler correctly identifies that the fidelity
gain is worth the modest queue time increase, which amounts to a few minutes even when Qurator's queue times are up to $5\times$ longer than
the least busy baseline. The least busy baseline pays average of 20--30 percentage
points of fidelity for this time saving, a poor tradeoff at low load. Qurator outperforms the fair scheduler, the list scheduler, at the same time producing more than 15\% better fidelity. The gains in queue times, especially at smaller circuit loads, are dominantly induced by selective deployment of circuit merging, which can be employed without significant penalties in the presence of smaller circuits and larger devices.
Fully equipped runs of Qurator show an improvement of over 50\% in queue times and a slight 2\% improvement in fidelity over Qurator runs with circuit transformations disabled on low load. The gap in queue time against the least busy baseline increases in medium qubit loads as circuit cutting imposes added load. The gap quickly closes as qubit counts increase. Qurator exhibits up to $30\%$ higher fidelity compared to QOS. For queue times under very light load of a handful of tasks, QOS starts ahead of Qurator by 50\% at the expense of this loss in fidelity, however Qurator quickly catches up at 10 tasks and beats QOS queue times by up to 70\% under low load while retaining its fidelity advantage. Fig. \ref{fig:splitloadpanelqod} shows the results for comparison against QOS. We only present results under low and medium load as QOS failed to terminate for high load conditions. In our experiments, we observe QOS taking up to 7x longer to make a decision compared to Qurator. 

As load grows into the medium regime, Qurator begins trading fidelity for
queue time. In the upper medium load range (3,000--5,000 tasks), Qurator
sacrifices up to 8\% fidelity relative to the highest-fidelity baseline
while reducing queue time by up to 75\%. Qurator offer fidelity as the number of qubits grow and present cutting opportunities. Compared to the least busy
baseline at the same load, Qurator accepts $3\times$ longer queue times
in exchange for 18\% higher fidelity. The fidelity loss at medium load is
acceptable in practice and can be compensated by increasing the number of
shots. At high load (5,000--35,000 tasks), the same pattern continues
with fidelity loss reaching 10\%. The small number of devices
(11) likely limits the scheduler's ability to find more favorable
tradeoffs; a larger device pool would be expected to improve results.  At medium load, Qurator retains a 50\% advantage in queue times over 
Qurator without transformations and the list scheduler, 60\% improvement over the fair share scheduler, with 20\% improvement in fidelity over all three baselines. Qurator retains its improvement in fidelity up to 18\%-20\% over QOS, while queue time savings exhibit up to 70\% gains. 

\paragraph*{High qubit results.} For high-qubit circuits, circuit cutting becomes dominantly applied to increase fidelity. While this increases the number of tasks in the system, which drives
up queue times and adds significant classical post-processing time, doubling the available device pool presents significant savings that dampen the effects of the associated overheads. The result is Qurator's commonly beating all baselines including least busy and highest fidelity baselines both on queue times and fidelity. This is explained by the fact that without circuit cutting, the baselines except QOS lose half of the available device pool due to lack of sufficient qubits. In the worst case observed, Qurator tracks least busy device closely with up to $11x$ longer queue times.  However, the least busy baseline achieves
less than 10\% probability of success for these circuits, making those
executions effectively worthless. Qurator's queue time overhead buys a
60\% fidelity gain over the list and fair share schedulers and up to a 54\% gain over the highest-fidelity device. On IBM devices specifically, Qurator leverages batch submission to reduce the queue overhead of cut circuits, achieving up to 25\% improvement in queue time over manually cut circuits submitted to the highest-fidelity device.

\paragraph*{Low variance conditions.} We observe a similarly important result when the variance in initial queue conditions is low, within at most a few hundred tasks of each other across all devices. We see Qurator beating both the highest fidelity device in terms of fidelity and the least busy device in terms of queue time at the same time by heuristic utilization of circuit transformations. This is again a significant result. While it was established that circuit cutting would increase execution fidelity, these results show that by increasing the device pool through enabling high fidelity execution on smaller devices, heuristic deployment of circuit cutting or merging can significantly reduce queue times and lead to better performance over the least busy accommodating device. This particularly makes Qurator a good fit for tightly coupled settings where this variance can be controlled and for the FTQC era where lower fidelity variance between devices can lead to more even demand.  

Qurator achieves a scheduling throughput of 0.4--1.5 tasks per second,
depending on the amount of cutting and merging required.

\paragraph*{Random load results.} Under random load conditions, Qurator
achieves on average a 3\% fidelity drop relative to the highest-fidelity
baseline with a 40\% reduction in queue time, and a 15\% fidelity gain
over the least busy baseline, with circuit merging disabled. Enabling
circuit merging yields a 60\% reduction in queue time at an average
fidelity cost of 12\%. We note that the MQT benchmark suite is heavily
dominated by smaller circuits, which limits Qurator's opportunity to
demonstrate the advantage of circuit cutting under random load conditions.

\section{Related Work}\label{sec:relatedwork}

Qurator sits at the intersection of  classical
heterogeneous scheduling, quantum cloud orchestration, and quantum network
scheduling. Prior work in each area has addressed parts of the problem, but
none has tackled the joint optimization of queue time and fidelity across
heterogeneous providers under the full set of quantum constraints. We survey
each area in turn and identify the specific gaps that Qurator fills.

\subsection{Classical Task Scheduling}

Classical scheduling has been studied extensively across single
processor~\cite{10.1145/3600006.3613163, 7013029, 10609594}, grid
systems~\cite{1039771, 765123}, data
centers~\cite{10.1145/3357223.3362728}, and
clusters~\cite{6217415, 1310764}, with algorithms spanning static list
scheduling~\cite{1303065, 765092, 993206, 10.1109/TPDS.2013.57,
radulescu2000fast}, task duplication~\cite{TANG2010323, 1264795},
genetic algorithms~\cite{OMARA201013, 7813746, 1420076, 6189509}, and
dynamic techniques~\cite{10.1137/0218016, 10.1016/0167-6377(88)90080-6,
10.5555/850940.852832, 6152724, 10.1145/3696355.3696361,
10.5555/2813767.2813803}. Qurator draws most directly from two
classical threads.

First, scheduling on shared cloud resources where the provider controls
access and exposes limited information~\cite{armstrong2010cloud, 7214080,
10.1145/2391229.2391254} closely mirrors the quantum cloud setting.
Sub-second scheduling~\cite{8705781} addresses the overhead bound imposed
by short-running tasks, a concern that applies directly to quantum circuits
that execute in under 10 seconds. Second, using historic runtime data to predict task behavior ~\cite{10.5555/646380.689540,
10.1145/2832105.2832109, historicgregg, 10.1145/3132211.3132451} is at the
foundation of Qurator.

However, three classical techniques that are widely used in heterogeneous
scheduling cannot be applied to quantum workloads. Preemptive and work
stealing schedulers~\cite{7013046, 8603197, 10.5555/3768039.3768055}
require the ability to suspend and migrate tasks, which quantum hardware
does not support. Task duplication is ruled out by the No-Cloning Theorem, restricted for tasks carrying unknown quantum state or entanglement.
Kernel slicing~\cite{6624111, 6582408}, which cuts GPU kernels into
sub-kernels for co-scheduling, has a quantum analogue in circuit cutting,
but quantum measurement collapse makes cutting fundamentally more expensive
than its classical counterpart. Various heterogeneous programming
models~\cite{10.1145/1346281.1346318, 10.1145/1383422.1383447, cuda,
openacc, opencl, 5375318, 10.1002/cpe.1631, condor} provide useful
abstractions that inspired Qurator's provider plugin architecture, but none
address quantum-specific constraints. While scheduling with
advance reservations~\cite{845974, 7152526, 4228032,
10.1145/2670979.2670981} is directly relevant to the quantum setting, its exploration remains out of the scope of this work given the limited number of providers offering reservations. 

\subsection{Quantum Orchestration and Cloud Scheduling}

The quantum software community has recognized the need for provider-agnostic
programming, producing frameworks \cite{Beisel2023QuantME4VQAMA,
seitz2023unifiedhybridhpcqctoolchain,
chundury2025scalinghybridquantumhpcapplications}, workflow
tools~\cite{10.1007/978-3-031-26507-5_35, electronics10080984},
conceptual architectures~\cite{10234288}, and platforms such as Amazon
Braket~\cite{awsbraket}. A unified toolkit for Quantum HPC integrating
quantum intermediate representations and control has also been
proposed~\cite{10821378}. However, these works either require users to
specify target providers explicitly, or perform hardware selection at a
primitive level based only on qubit count and gate set compatibility,
without considering queue time or fidelity jointly.

Several works extend classical HPC infrastructure to support quantum tasks.
SLURM extensions for quantum workflows have been proposed by multiple
groups, keeping SLURM’s fair-share semantic~\cite{slysz2025hybridclassicalquantumsupercomputingdemonstration,
chundury2025scalinghybridquantumhpcapplications, Shehata_2026, 10313739}.
Beck et al.~\cite{Beck_2024} propose an MPI-based task manager for tightly
integrated hybrid HPC with circuit cutting support. Shehata et
al.~\cite{Shehata_2026} propose a hardware-agnostic framework with
standardized resource management for interleaved hybrid workflows but do not make any heuristic scheduling decisions. 
Alvarado-Valiente et al.~\cite{ALVARADOVALIENTE2024103139} build a load
balancer for Amazon Braket, deploying standard fair-share scheduling in order to balance the workload without any consideration for fidelity impact or individual hardware semantics. Wild et al.~\cite{9233151} extend OpenTOSCA
for quantum orchestration. Li and Zhao~\cite{10707418} use reinforcement
learning for quantum serverless function orchestration. While these works
demonstrate efficient hybrid workflow execution, they seldom engage with the
unique properties of quantum devices that constrain scheduling decisions.

A smaller body of work investigates quantum-specific scheduling. While most of these systems assume one provider, one calibration schema, or one hardware family, Qurator targets the public cloud exposed to users: heterogeneous devices, inconsistent fidelity reports, incomplete topology exposure, limited queue visibility, and dynamic hybrid DAGs whose tasks may be generated at run time through circuit manipulation techniques. Additionally, these works often target isolated systems without any consideration for limitations imposed by the public cloud. Wang et al. \cite{Wang_2024} focus on scheduling VQE algorithms. They observe that not all training iterations require equally high fidelity and schedule earlier iterations on lower fidelity and lower queue devices. Seitz et
al.~\cite{Seitz_2024, Seitz_20242}, without considering fidelity properties, and Bhoumik et
al.~\cite{bhoumik2025distributedschedulingquantumcircuits} combine
scheduling with circuit cutting to maximize parallelism, but assume
dedicated QPU access and do not evaluate the impact of cutting on queue times. Their works focused solely on superconducting devices and their models commonly require job data that is not commonly exposed by today's providers. While these works use circuit cutting in scheduling context, they  deterministically deploy it to accommodate circuits that are too large to run on available devices and not as a heuristic selection tool. They do not consider the exponential impact of the postprocessing step or attempt automatic parameter optimization to find cuts that are favorable to the current resources. They require the tasks to be known statically prior to execution and do not fully support dynamic tasks. They only consider the individual task QPU execution length and don't consider the impact of queue times, initialization times or the added exponential overhead from the cut post-processing. Their heuristics are limited to the fitting on the QPU and not other important device characteristics. Their cut model only considers gate cutting, ignoring temporal cuts. Additionally, MILQ separates device mapping from scheduling. As we have shown in this paper, scheduling and decisions around circuit cutting are heavily intertwined with how the circuit maps on the device for cutting phase. Circuit merging has previously been used primarily to increase single-QPU utilization or group static tasks, rather than as a dynamic heuristic scheduling decision integrated with helper-qubit liveness reuse, hybrid DAG dependencies, and cross-device entanglement synchronization. Qonductor~\cite{giortamis2024orchestratingquantumcloudenvironments}
builds an end-to-end hybrid resource management platform on Kubernetes with a learning based estimator for fidelity and execution time. Their learning-based execution time estimator only predicts QPU runtime rather than public cloud queue wait times. Instead of an adaptive approach, they evenly distribute the jobs for higher cloud utilization. Their estimation of job semantics requires an active database of real job executions which is difficult to maintain and cannot adapt quickly to new devices without collecting the necessary training data first. They aim to maximize mean fidelity and minimize completion time across multiple tasks (avg. at 1000 tasks) and do not take individual task goals and characteristics into consideration. Although compatible with providers of different architectures, Qonductor is only concerned with gate-set compatibility and doesn’t take into consideration important hardware semantics we summarized in Section \ref{sec:fidest}. QGroup~\cite{10821381}
targets measurement synchronization and uses dynamic programming to group
circuits of similar length for parallel execution with the strong assumption of static tasks, specialized to superconducting devices. 

Most similar to Qurator,
Ravi et al.~\cite{ravi2022adaptivejobresourcemanagement} build an adaptive
scheduler that jointly optimizes fidelity and queue time on IBM devices,
using a model trained on historic data. Qurator builds on the core insight of Ravi et al. but targets a different operating regime. First, Ravi et al. instantiate their scheduler primarily for IBM superconducting devices, whereas Qurator canonicalizes calibration data across providers with different architectures, topologies, readout metrics, parallelization semantics (i.e. neutral-atoms) or analog abstractions. Scheduling across such architectures is more challenging than targeting a superconducting-device family with unified semantics. Second, their wait-time model relies on queue information such as exact queue times and execution estimates that is not generally exposed publicly. Qurator operates with weaker public signals, including queue length, calibration snapshots, and historical traces. Unlike its predecessors, which require large amounts of data to bootstrap, Qurator can quickly adapt to new devices through its small sliding-window approach. Third, Qurator treats cutting/merging as scheduler-level decisions rather than assuming a fixed circuit representation like Ravi et al. Fourth, Qurator models synchronized cross-device entangled task groups, where the objective includes start/finish skew within an entanglement lifetime rather than only DAG precedence. Finally, Ravi et al.'s fidelity estimation is loose, tracking average properties over the critical path. Previous works schedule individual quantum tasks, Qurator deploys end-to-end hybrid scheduling with consideration for dependencies and mandatory post-processing. The information Ravi et al. relied on is not readily accessible on today’s public quantum cloud, and the short operational lifetime of many devices, driven by rapid hardware turnover, further limits direct reproduction and controlled comparison. The devices they have tested on are long retired. 

Giortamis et al. introduce QOS \cite{giortamis2025qosquantumoperating}, a modular quantum operating system that combines error mitigation, fidelity estimation, multiprogramming, and scheduling to balance fidelity, waiting time, and resource utilization, evaluated on IBM QPUs. Qurator addresses a broader workflow scheduling problem by representing classical and quantum tasks within a common dynamic dependency graph, including tasks generated during execution. Its unified fidelity model reconciles heterogeneous calibration data and accounts for architecture-specific execution semantics, enabling scheduling across providers and hardware technologies. As opposed to QOS, Qurator takes device conditions including live, multi-provider device conditions and compilation-guided spatial and coherence-triggered temporal properties directly into consideration in its decision-making for circuit transformation. The impact of these can be seen in the comparison we present in Sec. \ref{sec:evaluation}. Qurator further adapts its fidelity v. queue-time tradeoff to observed load. These capabilities distinguish Qurator through integrated hybrid workflow scheduling, device dependent transformations and broader hardware coverage. 

In summary, these works are limited by one or more of the following: (1) They treat each quantum circuit as individual tasks and attempt to perform scheduling and orchestration over an independent set of entities, ignoring dependencies, synchronization barriers and classical pre/post processing costs. (2) They ignore the limitations brought on by the public cloud such as access limits, shared resources, lack of visibility into workload, availability concerns and inconsistent scheduling-relevant data. (3) Lack of joint optimization of queue times and fidelity. The works that attempt to optimize fidelity of execution face significant challenges in the public cloud. Based on the data we collected over a four month period, we observe high fidelity devices such as IBM Pittsburgh receive up to 1000x demand in terms of the number of circuits submitted. This congestion not only leads to significant wait time that can have crippling effects in hybrid workflows by blocking intensive classical tasks that depend on quantum layers, but also imposes a significant uncertainty in terms of fidelity. Long queue delays widen the interval between the calibration snapshot used for placement and the eventual execution time. Because device characteristics fluctuate over time, estimates based on stale calibration data may become less reliable. (4) Only integrating with the providers at an API level, without considering individual hardware semantics. (5) Adaptability. In today’s NISQ era, quantum hardware is rapidly evolving. Each iteration presents new semantics, different noise profiles, error-correction methods and access rules. Many of the existing works rely on large amounts of data collected from execution on these devices in order to make decisions. This presents a concern since today's quantum devices are frequently retired. In fact, our attempts to run these works have been unsuccessful in re-training their models or fitting predictive functions with the data available to us to a point where we can reproduce their reported results. For example, while Qonductor provided a pre-trained model as an artifact, this model was created two years ago and none of the devices it was trained on exist today. Therefore, we had to constrain our comparison to QOS, the sole existing work we could run and reproduce.

\subsection{Quantum Network Scheduling}

Scheduling entangled tasks introduces a dependency on quantum network
performance that has no classical analogue. Dahlberg et
al.~\cite{Dahlberg_2019} propose a link layer protocol for quantum networks
and consider basic scheduling in the context of entanglement distribution,
but do not study the scheduling problem in depth. To the best of our
knowledge, Qurator is the first scheduler to formally model and optimize
for the entanglement synchronization barrier in the context of a
multi-provider quantum cloud, and the benchmark metrics we introduce for
evaluating distributed entangled scheduling fill a gap in the literature. 

\section{Conclusion and Future Work}\label{sec:conc}

Quantum cloud scheduling is a fundamentally new problem. The 10--30,000$\times$
gap between circuit execution time and queue wait time makes queue-time
minimization the dominant concern, yet blindly chasing short queues destroys
fidelity. Classical scheduling techniques offer no remedy. Non-preemptibility,
the No-Cloning Theorem, dynamic DAG structure, heterogeneous gate sets, and
incompatible calibration data across providers each invalidate core scheduling
assumptions in ways that have no classical analogue.

Qurator addresses this by treating quantum constraints as first-class
scheduling concerns rather than workarounds. Circuit cutting and merging are
scheduling decisions, not programmer obligations. Fidelity estimation is
unified across six providers into a single logarithmic success score that
drives device selection alongside queue time. Entanglement synchronization
barriers are formalized and minimized as part of the scheduling plan. The
result is a scheduler that at low load matches the highest-fidelity baseline
within 1\%, and at high load achieves 30--75\% queue time reduction.

Our evaluation on four months of real queue data establishes, for the first
time, concrete baselines for what distributed entangled scheduling looks like
under real cloud conditions. The survival proxies we observed under realistic
queue lengths, ranging from $10^{-15}$ down to $10^{-239}$, make clear that
distributed quantum computing on today's public cloud is not yet viable. But
quantum networks are advancing rapidly, and the formal barrier model and
benchmark metrics we introduce here are designed to track that progress and
guide scheduler design as the landscape evolves.

Several directions remain open. \textbf{Advanced reservations}, now offered
by select providers such as Rigetti and IonQ, could substantially reduce
queue variance for entangled tasks, but the current one-hour reservation
granularity makes this impractical for short quantum circuits. As providers
reduce this granularity, reservation-aware scheduling becomes an important
next step. In select cases, \textbf{measurement
commutation}~\cite{11250243} can reduce circuit size without any postprocessing cost. Its impact on scheduling remains to be explored. \textbf{Dynamic rescheduling} is another natural extension: one
could cancel a job on one device and resubmit it elsewhere if a better
opportunity arises. However, queue length alone is insufficient to support
this, since a device with 50,000 small tasks may drain faster than one with
25,000 large tasks. Direct exposure of estimated wait times or per-job
characteristics from providers would be required. Additionally, we acknowledge that merging circuits for synchronized tasks results in a lower number of candidate devices. In cases where the queue length variance is high among devices, this could impact scheduling negatively. We omit a detailed exploration of this tradeoff in this paper and plan to address it in future work. 

\textbf{Error mitigation and fault-tolerant execution:} Qurator does not perform quantum error mitigation or
quantum error correction in the reported evaluation. We emphasize that error correction is also a scheduler level concern, since different devices will benefit from different error detecting codes and logical operator schemes, and the scheduler needs to distinguish error correction codes from core logic operations for an informed decision, further motivating the hybrid scheduling problem. We leave QEM support for future work. 

Finally, \textbf{reversibility and cross-compilation} present a longer-term
opportunity. In principle, any classical computation can be compiled to a
quantum circuit, and quantum circuits can be simulated classically, opening
new degrees of freedom in scheduling. However, for any workload a classical
device can handle, running it classically will almost always be more
efficient: quantum advantage is restricted to specific problem structures,
and today's quantum devices carry significant overhead in qubit count,
coherence constraints, and queue time. We do not pursue cross-compilation
here, but tracking the crossover point as quantum hardware improves is a
natural direction for future work. Most broadly, as fault-tolerant devices emerge
and coherence times grow, the tradeoff surface between fidelity, queue time,
and circuit cutting will shift, and Qurator's parametric design is intended
to adapt to that shift without architectural changes.

\textbf{FTQC-era: }
Qurator is evaluated for what is available to us: NISQ devices. However, the architecture is modular: the compiler, cutter, fidelity estimator, and provider adapters can be individually replaced. In an FTQC setting, physical-gate calibration terms would be replaced by logical-level quantities such as logical error rate, code distance, decoder latency, logical cycle time, magic-state factory availability, and networked Bell-pair generation rates, while core scheduler semantics remain intact. Some NISQ-specific decisions, especially cutting for physical-qubit limitations, could become less central. However, queue tradeoffs, heterogeneous selection, dynamic DAGs, non-preemptibility, and synchronization remain scheduling concerns. Balancing classical and quantum workloads remains central to Qurator. Dynamic delegation to classical resources based on both device and factory availability constraints, and cost concerns are expected to be central to the FTQC era. This makes Qurator's core semantics valuable beyond the NISQ era. 

\section*{Author Contributions}

S.P. conceived the idea, developed the fidelity framework, cutting and merging semantics, entanglement barriers and the overall scheduler. U.d.M. developed and tested the CutQC \cite{tangCutQCUsingSmall2021} integration. A.S. and P.K. provided direction. All authors discussed the results and contributed to
writing the manuscript. LLMs were used to create the outline for Fig. 3, but the contents were created by the authors. LLMs were used to proofread the manuscript, any recommendations were manually evaluated and addressed by the authors. 

\bibliographystyle{ACM-Reference-Format}
\bibliography{ref}
\end{document}